\documentclass[twocolumn,amsmath, aps, amssymb, superscriptaddress]{revtex4}
\usepackage{dcolumn}% Align table columns on decimal point
\usepackage{bm}% bold math
\usepackage{amsmath}
\usepackage{amsfonts}
\usepackage{graphics}
\usepackage{graphicx} 
\usepackage{verbatim}
\usepackage{enumerate}
\usepackage{color} 

\begin{document}

\title{Persistence and Uncertainty in the Academic Career}
\author{Alexander M. Petersen}
\affiliation{Laboratory for the Analysis of Complex Economic Systems, Institutions Markets Technologies (IMT) Lucca Institute for Advanced Studies, Lucca 55100, Italy}
%\affiliation{Center for Polymer Studies and Department of Physics, Boston University, Boston, Massachusetts 02215, USA}
\author{Massimo Riccaboni}
\affiliation{Laboratory of Innovation Management and Economics, IMT Lucca Institute for Advanced Studies, Lucca 55100, Italy}
\affiliation{Crisis Lab, IMT Lucca Institute for Advanced Studies, Lucca 55100, Italy}
\author{H. Eugene Stanley}
\affiliation{Center for Polymer Studies and Department of Physics, Boston University, Boston, Massachusetts 02215, USA}
\author{Fabio Pammolli}
\affiliation{Laboratory for the Analysis of Complex Economic Systems, Institutions Markets Technologies (IMT) Lucca Institute for Advanced Studies, Lucca 55100, Italy}
\affiliation{Laboratory of Innovation Management and Economics, IMT Lucca Institute for Advanced Studies, Lucca 55100, Italy}
\affiliation{Crisis Lab, IMT Lucca Institute for Advanced Studies, Lucca 55100, Italy}
\affiliation{Center for Polymer Studies and Department of Physics, Boston University, Boston, Massachusetts 02215, USA}
\begin{abstract} 
Understanding how institutional  changes within academia 
may affect the overall potential of science requires a better quantitative representation of how careers evolve over
time. Since knowledge spillovers, cumulative advantage, competition, and collaboration are distinctive features of the academic
profession, both the employment relationship and the procedures for assigning 
recognition and allocating funding  should 
be designed to account for these factors. We study the annual production $n_{i}(t)$ of a 
given scientist  $i$ by analyzing  longitudinal career data for 200 leading scientists and 100 assistant professors from the physics community. 
We compare our results with  21,156 sports careers. 
Our empirical analysis of individual productivity dynamics shows that (i) there are increasing returns for the top 
individuals within the competitive cohort, and that (ii) the distribution of production growth is a leptokurtic ``tent-shaped'' distribution that is remarkably symmetric.
Our methodology is general, and we speculate that similar features appear in other disciplines where academic publication is essential and collaboration is a key feature.
We introduce a model of proportional growth which reproduces these two observations, and additionally accounts for the significantly right-skewed  distributions of career longevity and achievement in science.
Using this theoretical model, we show that short-term contracts can amplify the effects of competition and uncertainty making careers more vulnerable to  early termination, not necessarily due to lack of individual talent and persistence, but because of random negative production shocks.
We show that fluctuations in  scientific  production are quantitatively related to 
a scientist's collaboration radius and team efficiency.    
\end{abstract}
\date{\today}

%\pacs{01.80.+b,89.75.Da,02.50.Fz}
\maketitle
%\section{Introduction}
\footnotetext[1]{ Corresponding author: Alexander M. Petersen \newline
\emph{E-mail}: petersen.xander@gmail.com \newline or alexander.petersen@imtlucca.it}

Institutional change could alter the relationship between Science and scientists as well as the
 longstanding  patronage system in academia  \cite{OpenScience, InnovAcc}. 
Some recent  shifts in academia  include the changing business structure of research universities \cite{TenureBook}, shifts in the  labor supply-demand balance \cite{ScienceSTEM}, a bottleneck 
in the number of  tenure track positions \cite{PhdFactory},  and a related policy shift away from long-term contracts \cite{TenureBook,Tenure}. 
Along these lines, significant factors for consideration  are the increasing range in research team size \cite{TeamScience}, the 
economic organization required to fund and review  collaborative research projects, and  the evolving definition of 
the role of the academic research professor \cite{TenureBook}.

The role of individual performance metrics in career appraisal, in domains as diverse as sports \cite{BB1,BB3}, finance \cite{FinanceSteroids, FinanceIM}, and academia, is increasing in this data rich age. 
In the case of academia, as the typical size of  scientific collaborations  increases \cite{TeamScience}, the allocation of  funding and the
association of
recognition at the varying scales of science (individual $\leftrightarrows$ group $\leftrightarrows$ institution \cite{MultilevelScience}) has become more complex. Indeed, scientific achievement is becoming increasingly linked to online visibility in a considerable reputation tournament  \cite{Popularity}. 

Here we seek to  identify (i)  quantitative patterns in the scientific career trajectory towards a better understanding of career dynamics and  achievement
\cite{ShockleyProductivity,careertrajectory,BB2,gappaper,SciCreditDiffusion, Scientists,AgeDynamicsNobel}, and  (ii)  
how  scientific production responds to policies concerning contract length.  
Using rich productivity data available at the level of single individuals, we analyze longitudinal career data keeping in mind the
the roles of spillovers, group size, and career sustainability. 
Although our empirical analysis is limited to careers in physics, our approach is general. We speculate 
that similar features describe other disciplines where academic publication is a primary indicator and collaboration is a key feature. 

Specifically, we analyze production data 
 for 300 physicists  $i=1...300$
who are distributed into 3 groups: (a) Group A corresponds to the 100 most cited physicists with average $h$-index
$\langle h \rangle = 61 \pm 21$, (b) Group B corresponds to 100 additional highly-cited physicists with 
 $\langle h \rangle = 44 \pm 15$, and (c) Group C corresponds to 100  assistant professors in 50 U.S. physics
departments with  $\langle h \rangle = 15 \pm 7$. 
We  define the annual production 
$n_{i}(t)$ as the number of papers published by scientist $i$ in year $t$ of his/her career. We focus on academic careers from the physics community to
approximately control for significant cross-disciplinary production variations. Using the same set of scientists, a  companion study has analyzed the 
rank-ordered citation distribution of each scientist with a focus on the statistical regularities underlying
publication impact  \cite{gappaper}. 
%We balance our longitudinal study  by  analyzing empirical panel data for a non-academic labor force  comprising 21,156 sports careers from  two
%prestigious American sports leagues: the set of all Major League Baseball (MLB) careers during the 90-year period 1920--2009 and
% National Basketball Association (NBA) careers during the 63-year period 1946--2008. 
%For non-intellectual labor we define $n_{i}(t)$ 
%using in-game opportunity and success measures.
 %While these two professions both display a high level of competition, they differ in their employment
%term structure and salary scale.  In the case of academia, the  tenure system rewards high performance levels with
% lifelong employment (tenure). In contrast,  professional sports are characterized by relatively short contracts that
%emphasize continued performance over a shorter time frame and thereby exploit the high levels of athletic prowess in a player's peak years.
 We provide further description of the data and present a parallel analysis of 21,156 sports careers
in the Supporting Information Appendix (SI) text.

We begin this paper with empirical analysis of longitudinal career data. Our empirical evidences serve as statistical benchmarks used in the final section where we develop a stochastic proportional growth model. 
In particular, our model shows that a short-term appraisal system can  result in a significant number of  ``sudden'' early deaths due to unavoidable
negative production shocks.  This result is consistent with a Matthew Effect model \cite{BB2} and recent academic career survival analysis \cite{survivalanalysis}, which 
demonstrate how young careers can be stymied by the difficulty in overcoming early achievement barriers. Altogether, our results indicate that short-term contracts may increase the strength of the
``rich-get-richer'' mechanism in science \cite{Matthew1,DeSollaPrice} and  may hinder the upward mobility of young scientists.

\section{Results}
\subsection{Scientific production and the career trajectory} \
The academic career depends on many factors,  such as cumulative advantage
\cite{BB2, Scientists,Matthew1,DeSollaPrice}, the ``sacred spark,'' 
\cite{GrowthDynamicsH,ProdDifferences}, and other complex aspects of knowledge transfer manifest in our techno-social world \cite{TechnoSocial}. 
To exemplify this complexity, a recent case study on the impact trajectories of Nobel prize winners shows that ``scientific career
shocks'' marked by the publication of an individual's ``magnum opus'' work(s) can trigger future recognition and reward,
resembling the cascading dynamics of earthquakes \cite{citationboosts}. 

\begin{figure}[t]
\centering{\includegraphics[width=0.45\textwidth]{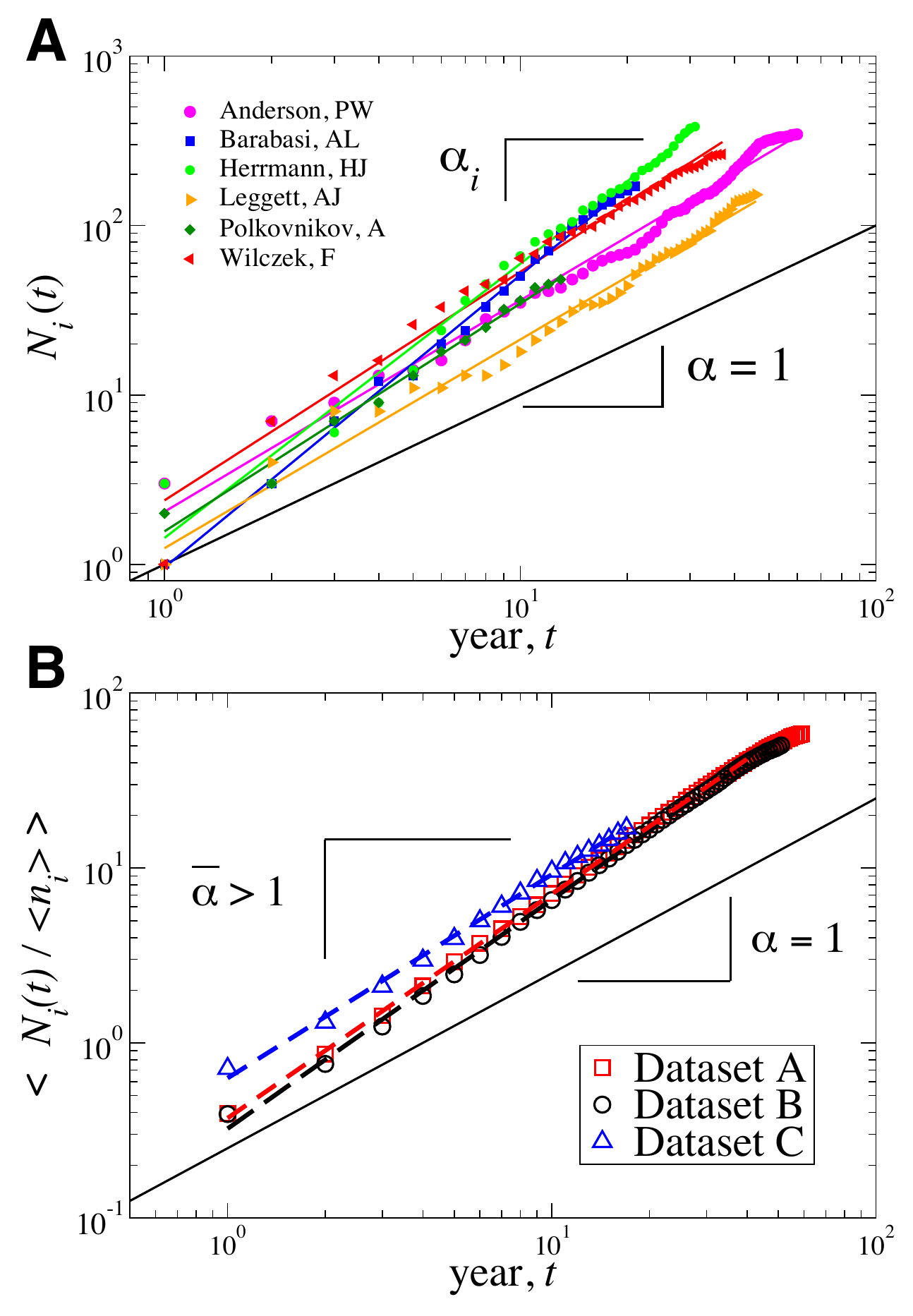}}
\caption{ Persistent accelerating career growth. (A) The career trajectory $N_{i}(t) \sim t^{\alpha_{i}}$ of
six stellar careers from varying age cohorts.  The $\alpha_{i}$ value characterizes the career persistence, where 
careers with $\alpha>1$ are accelerating. $\alpha_{i}$ values calculated using OLS regression in alphabetical order are:
%$\alpha =1.05\pm0.01$ (K. Malone), $\alpha = 1.05 \pm0.01$ (P. Rose), 
$\alpha = 1.25\pm 0.02$, 
%(Anderson, PW), 
$\alpha = 1.72\pm 0.02$, 
%(Barabasi, AL),
$\alpha = 1.62\pm 0.04$, 
%(Herrmann, HJ), 
$\alpha = 1.23\pm 0.02$, 
%(Leggett, AJ),
$\alpha = 1.34\pm 0.05$, 
%(Polkovnikov, A), and 
$\alpha = 1.35\pm 0.04$. 
%(Wilczek, F). 
(B) Defined in Eq. [\ref{Npt}], the average career trajectory $\langle N'(t) \rangle$ calculated from 100 individual  $N_{i}(t)$  in each
dataset demonstrates robust accelerating career growth within each cohort. We use the normalized career trajectory $N'_{i}(t)$  in order to aggregate $N_{i}(t)$ with varying publication rates $
\langle n_{i} \rangle$. As a result, the aggregate scaling exponent $\overline \alpha$ quantifies the acceleration of
the typical career over time, independent of $\langle n_{i} \rangle$.    For the scientific careers, we calculate 
$\overline \alpha$ values: $1.28\pm0.01$  [A], $1.31\pm0.01$ [B], and $1.15\pm0.02$ [C].
   These values are all significantly greater than unity, $\overline \alpha >1$, indicating that cumulative advantage in science is closely related to knowledge and production
spillovers. We calculate $\overline \alpha$ using OLS regression and
plot the corresponding best-fit lines (dashed) for each dataset. \label{NTi}}
\end{figure}

We model the  career trajectory as a sequence of scientific outputs which arrive at the variable rate  $n_{i}(t)$. Since the
reputation of a scientist is typically  a cumulative representation of his/her contributions, we consider the cumulative
production $N_{i}(t) \equiv
\sum_{t'=1}^{t}n_{i}(t')$ as a proxy for career achievement. Fig. \ref{NTi}(A) shows the cumulative production $N_{i}(t)$
 of six notable careers which display a temporal scaling relation 
$N_{i}(t) \approx A_{i} t^{\alpha_{i}}$ where $\alpha_{i}$ is a scaling exponent that quantifies the career trajectory dynamics. 
  The average and standard deviation of the $\alpha_{i}$ values calculated for  each dataset are $\langle \alpha_{i}
\rangle =  1.42 \pm 0.29$  [A], $1.44 \pm 0.26$ [B], and $1.30 \pm 0.31$ [C].  
We justify this 2-parameter model  in the SI Appendix text  using scaling methods and data collapse.

There are also numerous cases of $N_{i}(t)$ which do
not exhibit such 
regularity (see Fig. S1), but instead display marked non-stationarity and non-linearity arising from significant exogenous career
shocks. Positive  shocks, possibly corresponding to just a single discovery, can spur significant  productivity and reputation growth \cite{GrowthDynamicsH,citationboosts}. 
Negative  shocks, such as in the case of scientific fraud, can end the career rather suddenly. We also acknowledge that
the end of the career is a difficult phase to analyze, since such an event can occur quite abruptly, and so our analysis is mainly concerned with the growth phase and 
not the termination phase. 

%Fig1 here

In order to analyze the average properties of $N_{i}(t)$ for all 300 scientists in our sample, we define the normalized
trajectory $N'_{i}(t) \equiv N_{i}(t) / \langle n_{i} \rangle$. The quantity $\langle n_{i} \rangle$ is the average annual
production of author $i$, with $N'_{i}(L_{i}) = L_{i}$ by construction ($L_{i}$ corresponds to the career length of individual $i$). 
Fig. \ref{NTi}(B) shows the characteristic production trajectory obtained by averaging together the 100  $N'_{i}(t)$ belonging to
each dataset, 
\begin{equation}
\langle N'(t) \rangle \equiv \Big \langle \frac{N_{i}(t)}{ \langle n_{i} \rangle} \Big \rangle  \equiv \frac{1}{100}
\sum_{i=1}^{100} \frac{N_{i}(t)}{ \langle n_{i} \rangle} \ .
\label{Npt}
\end{equation}
The standard deviation $\sigma(N'(t))$  shown in Fig. S2(B) begins  to decrease after roughly 20 years for dataset [A] and [B] scientists.
Over this horizon, the stochastic arrival of career shocks  can significantly alter the career trajectory \cite{AgeDynamicsNobel,GrowthDynamicsH,citationboosts,StarDeath}. 
Each $N'_{i}(t)$  exhibits robust scaling corresponding to the  scaling law $\langle N'(t) \rangle \sim t^{\overline \alpha}$.
This regularity reflects the abundance of of careers with $\alpha_{i} >1$  corresponding to accelerated career growth. This acceleration is consistent with increasing returns  arising from knowledge and production spillovers.

\subsection{Fluctuations in scientific output over the academic career} 
\begin{figure}
\centering{\includegraphics[width=0.49\textwidth]{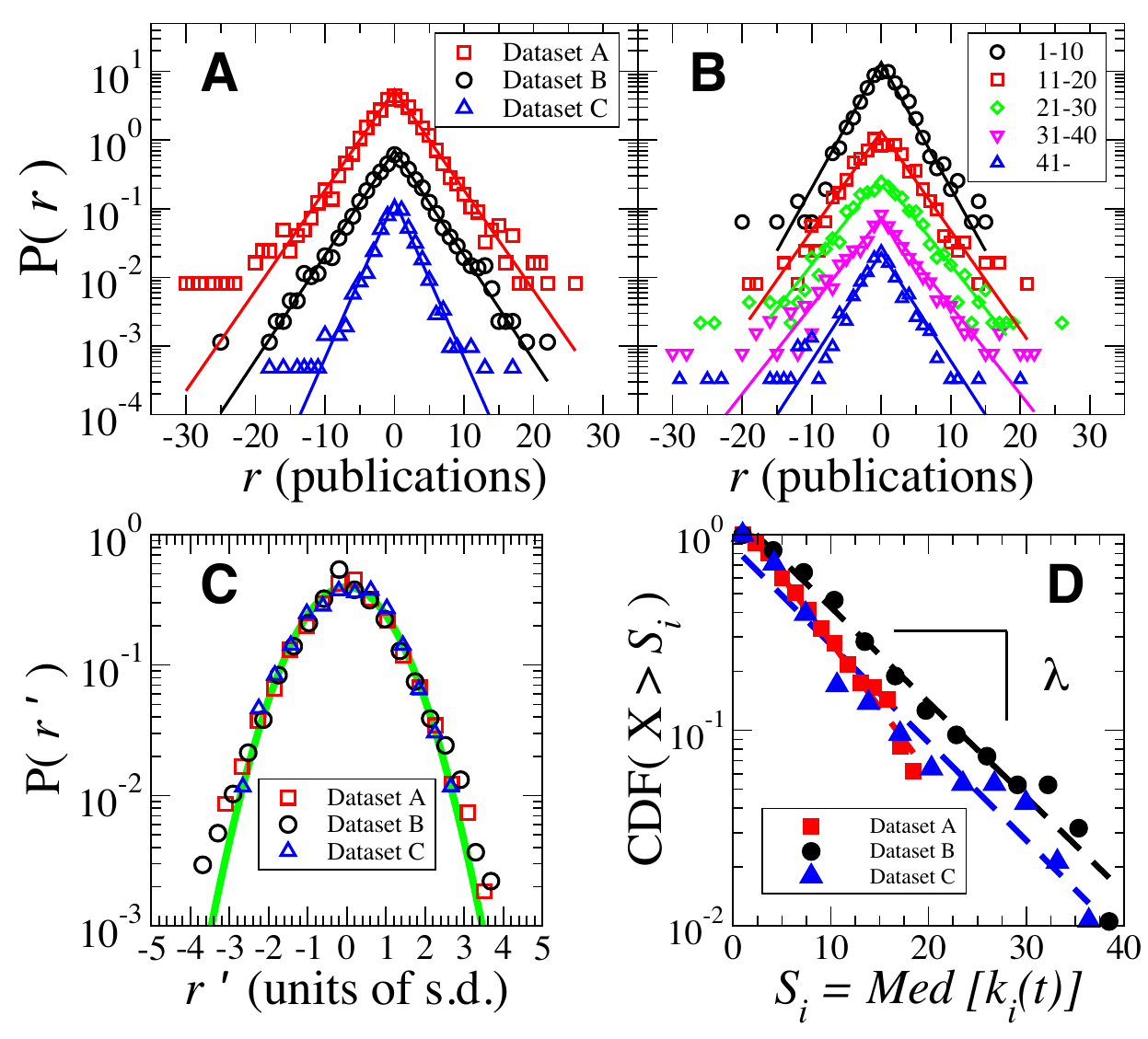}}
\caption{ Empirical evidence for the  proportional growth model of career production.
%Consider the career of Michael Jordan, who suffered a broken foot during his second professional NBA season.
%Nevertheless, with the confidence of his employer, MJ reestablished his  investment value through persistence and
%resilience. The following year MJ achieved an extremely large $R_{i}(t) = 7.6  \sigma$ growth value. 
  (A) Probability density function (pdf) of the annual production change $r$ in the number of papers published over
a $\Delta t =1$ year period. 
 In the bulk of each $P(r)$, the growth distribution is approximately double-exponential (Laplace).  
(B) To test the stability of the distribution over career trajectory subintervals, we separate $r_{i}(t)$ values
into 5 non-overlapping 10-year periods and verify the stability of the Laplace $P(r)$. For each $P(r)$, we also plot the
corresponding 
Laplace distribution (solid line) with standard deviation $\sigma$ and mean $\mu \approx 0$ calculated using the maximum
likelihood estimator method.  To improve graphical clarity, we vertically offset  each $P(r)$ by a constant factor.   
 For visual comparison, we also plot a Normal distribution (dashed black curve) with $\sigma \equiv 1$
which instead decays parabolically on the log-linear axes.
(C) Accounting for individual production factors by using the normalized production change $r'$, the resulting pdfs $P(r')$ 
collapse onto a Gaussian distribution with unit variance.
 Deviations in the tails likely correspond to extreme ``career shocks.'' (D)
 The cumulative distribution  $CDF(X\geq S_{i})$ is exponential, indicating that the unconditional distributions $P(r)$ in (A) and (B) follow from an 
 exponential mixing of conditional Gaussian distributions $P(r|S_{i})$.   \label{dNPDF}} 
\end{figure}

\
Individuals are constantly entering and exiting the professional market, with birth and death rates depending on complex
economic and institutional factors. Due to competition,  decisions and performance at the early stages of the career can  have long
lasting consequences   \cite{BB2,Entrance}.   
To better understand career uncertainty portrayed by the common saying ``publish or perish" \cite{OutputRecognition}, we analyze the outcome
fluctuation 
\begin{equation}
r_{i}(t) \equiv  n_{i}(t)-n_{i}(t-\Delta t) \ 
\label{Dn}
\end{equation}
of career $i$ in year $t$ over the time interval $\Delta t=1$ year. Fig. \ref{dNPDF}(A) and (B) show the 
unconditional pdf of $r$ values which are leptokurtic but remarkably symmetric, illustrating 
the endogenous frequencies of positive and negative output growth. Output fluctuations arise naturally from the  lulls
and bursts in both the mental and physical capabilities of humans \cite{Barabasibursts,InvPercHumanDyn}. 
Moreover, the statistical regularities in the annual production change distribution indicate a
striking resemblance to the growth rate distribution of countries, firms, and universities \cite{Growth12,Growth6}.

%Fig 2 Here

To better account for individual growth factors, we next define  the normalized production change
\begin{equation}
r'_{i}(t) \equiv  [r_{i}(t)- \langle r_{i} \rangle ] / \sigma_{i}(r) 
\label{rnorm}
\end{equation}
which is measured in units of the fluctuation scale  $\sigma_{i}(r)$ unique to each career.
We measure the average $\langle r_{i} \rangle$ and the standard deviation $\sigma_{i}(r)$ of each  career   using the first  $L_{i}$ available years for
each scientist $i$. $r'_{i}(t)$ is a better measure for comparing career uncertainty, 
since individuals have production factors that depend on the type of research, the size of the collaboration team, and
the position within the team. 
 Fig. \ref{dNPDF}(C) show that $P(r')$, the  probability density function (pdf) of $r'$ measured in units of standard deviation,  is well approximated by a Gaussian
distribution with unit variance. The data collapse of each $P(r')$ onto the predicted Gaussian distribution  (solid green curve) indicates that 
individual output fluctuations are consistent with a proportional growth model. We note that the remaining deviations in the tails for $\vert r' \vert \geq 3$ are likely signatures of the
exogenous career shocks that are not accounted for by an endogenous proportional growth model.

The ability to collaborate on large projects, both in close working teams and  in extreme examples as remote agents (i.e. Wikipedia \cite{Collaborationsociety}), is  one of the
foremost properties of human society.
In science, the ability to attract future opportunities is strongly related to  production and knowledge
spillovers   \cite{StarDeath, Spillovers1, Spillovers2} that are facilitated by  the collaboration network
\cite{TeamScience,MultilevelScience,Borner,socialgroupevol,CoauthorshipNetworks,ArxivNetwork,TeamAssembly}. 
Indeed, there is a tipping point in a scientific career that
occurs when a scientist's knowledge investment reaches a critical mass that can sustain production over a long horizon,  and when a scientist becomes an attractor (as opposed to a pursuer) of 
new collaboration/production opportunities. To  account for
 collaboration, we calculate for each author the number $k_{i}(t)$ of distinct coauthors per
year and then define his/her collaboration radius $S_{i}$ as the median of the set of his/her $k_{i}(t)$ values, $S_{i} \equiv Med[k_{i}(t)]$. 
We use the median instead of the average $\langle k_{i}(t) \rangle$ since extremely large 
$k_{i}(t)$  values can occur in specific  fields such as high-energy physics and astronomy. 

 %Fig 3 here
\begin{figure}
\centering{\includegraphics[width=0.45\textwidth]{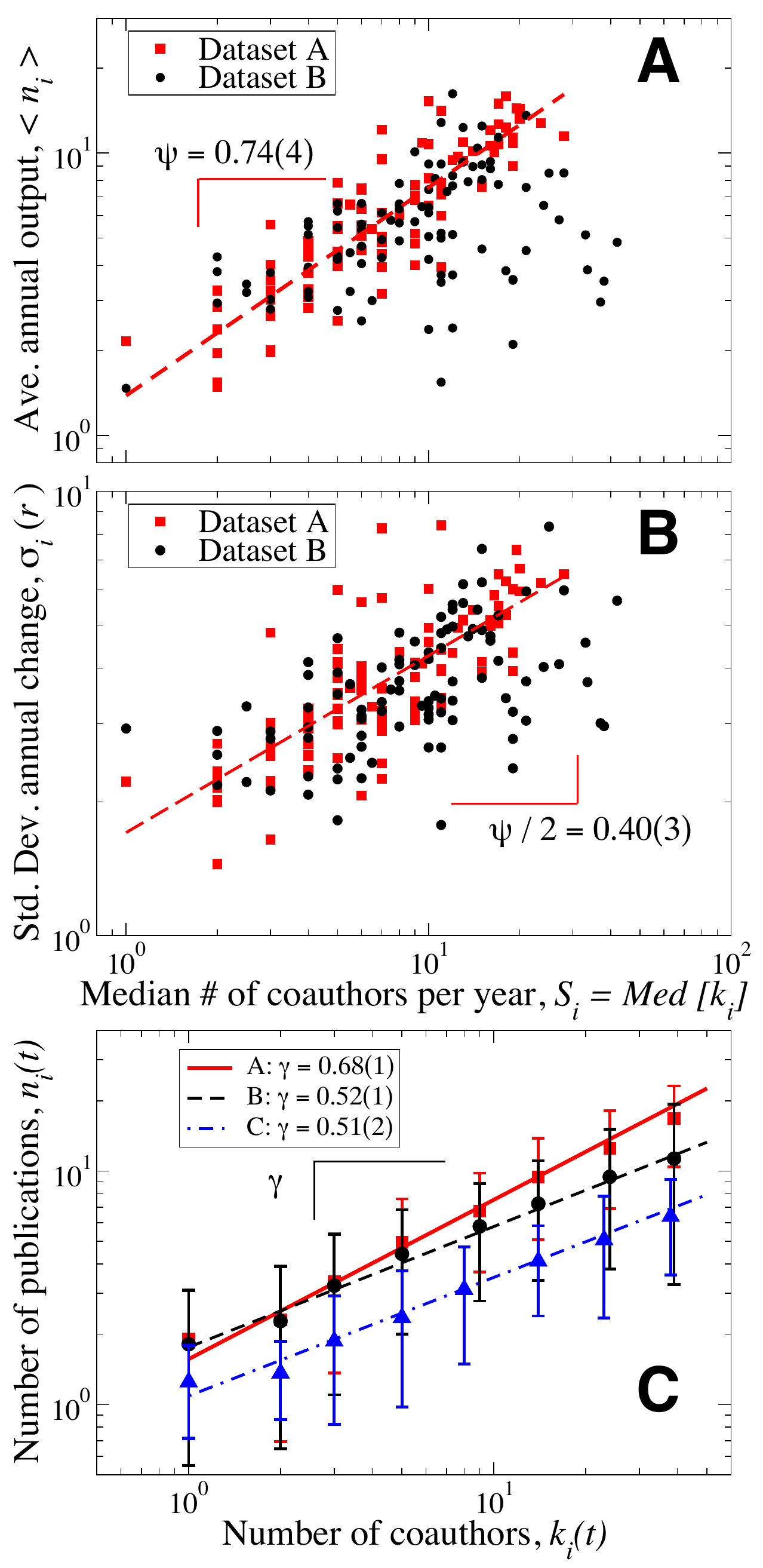}}
 \caption{ Quantitative relations between career growth, career risk,  and collaboration efficiency.
The fluctuations in production reflect the 
 unpredictable horizon of ``career shocks'' which can affect  the ability of a scientists to access new 
creative opportunities.
 (A) Relation between average annual production  $\langle n_{i} \rangle$ and collaboration radius $S_{i}
\equiv Med[ k_{i}]$ shows a decreasing marginal output  per collaborator as demonstrated by sublinear $\psi<1$. Interestingly,
dataset [A] scientists have on average a larger output-to-input efficiency.
 (B) The production fluctuation scale $\sigma_{i}(r) $ is a quantitative measure for uncertainty in academic
careers, with
 scaling relation $\sigma_{i}(r) \sim S_{i}^{\psi / 2}$. 
   (C)  Management, coordination, and training inefficiencies can result in a $\gamma <1$ 
 corresponding to a decreasing marginal return with each additional coauthor input. The significantly larger $\gamma$ value for 
 dataset [A] scientists seems to suggest that managerial abilities related to output efficiency is a common attribute of top scientists.
\label{NSDscaling}} 
\end{figure}

Given the complex scientific coauthorship network, we ask the question: what is  the typical number of unique coauthors per year? 
Fig. \ref{dNPDF}(D) shows the  cumulative distribution function $CDF(S_{i})$ of $S_{i}$ values for each data set. The approximately linear form on log-linear axes indicates  that  $S_{i}$ is exponentially distributed, $P(S_{i}) \sim \exp[-\lambda S_{i}]$. 
We calculate  $\lambda = 0.15 \pm 0.01$ [A], 
$\lambda = 0.11\pm 0.01$ [B], and $\lambda = 0.11 \pm 0.01$ [C]. 
The exponential size distribution has
been shown to emerge in complex systems 
where linear preferential attachment  governs the acquisition of new opportunities \cite{ExpDistSize}.
This result shows that 
the leptokurtic ``tent-shaped''   distribution $P(r)$ in Fig. \ref{dNPDF}  
follows from the exponential mixing of heterogenous conditional Gaussian distributions  \cite{MixingGaussians}. 

The exponential mixture of Gaussians decomposes the unconditional distribution $P(r)$ into  a mixture of conditional  Gaussian distributions 
\begin{equation} 
%P(r| S) = \exp[ -r^{2}/2\sigma^{2}(r)] / \sqrt{2\pi \sigma^{2}(r)} \ .
P(r| S_{i}) = \exp[ -r^{2}/2VS_{i}^{\psi}] / \sqrt{2\pi VS_{i}^{\psi}} \ ,
\end{equation}
each with a fluctuation scale $\sigma_{i}(r)$ depending on $S_{i}$ by the scaling relation
\begin{equation}
\sigma^{2}_{i}(r) \approx V S_{i}^{\psi} \ .
\label{PsiScaling}
\end{equation}
Hence, the mixture is  parameterized by $\psi$
\begin{equation} 
P_{\psi}(r) = \int_{0}^{\infty}  P(r|S) P(S) dS  \approx \sum_{i=1}P_{i}(r|S_{i}) P(S_{i}) \ .
\label{expmix0}
\end{equation}
The independent case $\psi =0$ results in a Gaussian $P_{\psi}(r)$ and the linear case $\psi =1$ results in a Laplace (double-exponential) $P_{\psi}(r)$.
See the SI Appendix text and ref. \cite{MixingGaussians} for further discussion of the $\psi$ dependence of $P_{\psi}(r)$.

\subsection{The  size-variance relation and group efficiency}
The values of $\psi$ for scientific  and athletic careers follow from the different combination of physical and
intellectual inputs that enter the production function for the two distinct professions.  Academic knowledge is
typically a non-rival good, and so knowledge-intensive professions  are characterized by spillovers, both over time and
across collaborations \cite{Spillovers1,Spillovers2}, consistent with $\alpha_{i}>1$ and $\psi >0$.  Interestingly, Azoulay et al. show evidence for production  spillovers  in the 5--8\% decrease in  output by scientists who were close collaborators with a ``superstar''
scientists who died suddenly \cite{StarDeath}.

We now formalize the quantitative link between scientific collaboration  \cite{Borner,socialgroupevol} and career growth given by the
size-variance  scaling relation in Eq.~[\ref{PsiScaling}] visualized in the scatter plot in Fig. \ref{NSDscaling}(B). Using ordinary least squares (OLS) regression of the data on log-log scale, we calculate  $\psi/2 \approx 0.40\pm 0.03$ ($R =0.77$) for dataset [A], $\psi/2 \approx 0.22 \pm
0.04$ ($R = 0.51$) [B], and $\psi/2 \approx 0.26 \pm 0.05$ ($R =0.45$) [C]. 
Interdependent tasks characteristic of group collaborations typically involve partially overlapping efforts.
Hence, the empirical $\psi$ values are significantly less than the value $\psi = 1$ that one would expect from the sum of $S_{i}$ independent random variables with approximately equal variance $V$.   
Collectively, these empirical evidences serve as coherent motivations for the the preferential capture  growth model that we propose in the following section.

Alternatively, it is also possible to estimate $\psi$ using the relation  between the average annual production $\langle n_{i} \rangle$ and the collaboration radius $S_{i}$.
The input-output  relation $\langle n_{i} \rangle \sim S_{i}^{\psi}$ quantifies the collaboration efficiency, with $\psi = 0.74 \pm 0.04$ ($R=0.87$) for  dataset [A]  and $\psi = 0.25 \pm 0.04$ ($R=0.37$) for dataset [B]. 
If the autocorrelation between sequential production values
$n_{i}(t)$ and $n_{i}(t+1)$  is relatively small, then we expect the scaling exponents calculated for $\langle n_{i} \rangle$ and $\sigma^{2}_{i}(r)$ to be 
approximately equal. This result follows from considering $r_{i}(t)$ as the convolution of an underlying production distribution 
$P_{i}(n)$ for each scientist that is approximately stable.  Interestingly,  the larger $\psi$ values calculated   for dataset [A] scientists
suggests that prestige is related to the  increasing  returns
 in the scientific production function \cite{IncreasingReturns}.

Next we use an alternative method to estimate the annual collaboration efficiency by relating the number of publications  $n_{i}(t)$  in a given year to the  number of distinct coauthors $k_{i}(t)$ over the same year.  
We use a  single-factor  production function, 
\begin{equation}
n_{i}(t) \approx q_{i}[k_{i}(t)]^{\gamma_{i}} \ ,
\label{gammaefficiency}
\end{equation}
to quantify the relation between output and  labor inputs with a scaling exponent
$\gamma_{i}$. We estimate $q_{i}$ and $\gamma_{i}$ for each author using OLS regression, and define the normalized output measure $Q_{i} \propto n_{i}(t)/k_{i}(t)^{\gamma_{i}}$ 
using the best-fit $q_{i}$ and $\gamma_{i}$ values calculated for each scientist $i$.  Fig. \ref{NSDscaling}(C) shows  the  efficiency parameter $\gamma$ calculated by aggregating all careers in each dataset, and indicates that this aggregate $\gamma$ is approximately equal 
to the average $ \langle \gamma_{i}\rangle$ calculated from  the  $\gamma_{i}$ values in each career dataset:
 $\gamma = 0.68 \pm 0.01$ [A], $\gamma = 0.52 \pm 0.01$ [B], and $\gamma = 0.51 \pm 0.02$ [C].
 Furthermore, the   $\psi$ and $\gamma$ values are approximately equal, which is not surprising, since both scaling exponents are efficiency 
 measures that relate the scaling relation of output $n_{i}(t)$ per input $k_{i}(t)$.

\subsection{A Proportional growth  model for scientific output} \ We develop a stochastic model  as a heuristic tool to better understand the 
effects of long-term versus short-term contracts.  In this  competition model,   opportunities (i.e. new scientific publications) are captured according to
a general mechanism whereby the  capture rate $\mathcal{P}_{i}(t)$ depends on the appraisal $w_{i}(t)$ of an individual's 
record of achievement over a prescribed history. We define the appraisal to be an exponentially weighted average over a given individual's history of production
\begin{equation}
w_{i}(t) \equiv \sum_{\Delta t =1}^{t-1} n_{i}(t-\Delta t) e^{-c\Delta t} \ ,
\label{wit}
\end{equation}
which is characterized by the appraisal horizon $1/c$. 
 We use the value $c=0$  to represent a long-term appraisal (tenure) system and a value $c \gg 1$ to represent a short-term appraisal system. Each agent $i=1...I$ simultaneously attracts new opportunities at a rate
\begin{equation}
\mathcal{P}_{i}(t) = \frac{w_{i}(t)^{\pi}}{\sum_{i=1}^{I} w_{i}(t)^{\pi}} \ .
\label{Pi}
\end{equation}
until all $P$  opportunities for a given period $t$ are captured. We assume that each agent has the production potential of one unit per period, and so the total number of opportunities distributed per period $P$ is equal to the number of competing agents, $P\equiv I$. 

%Fig 4 here
\begin{figure*}
\centering{\includegraphics[width=0.98\textwidth]{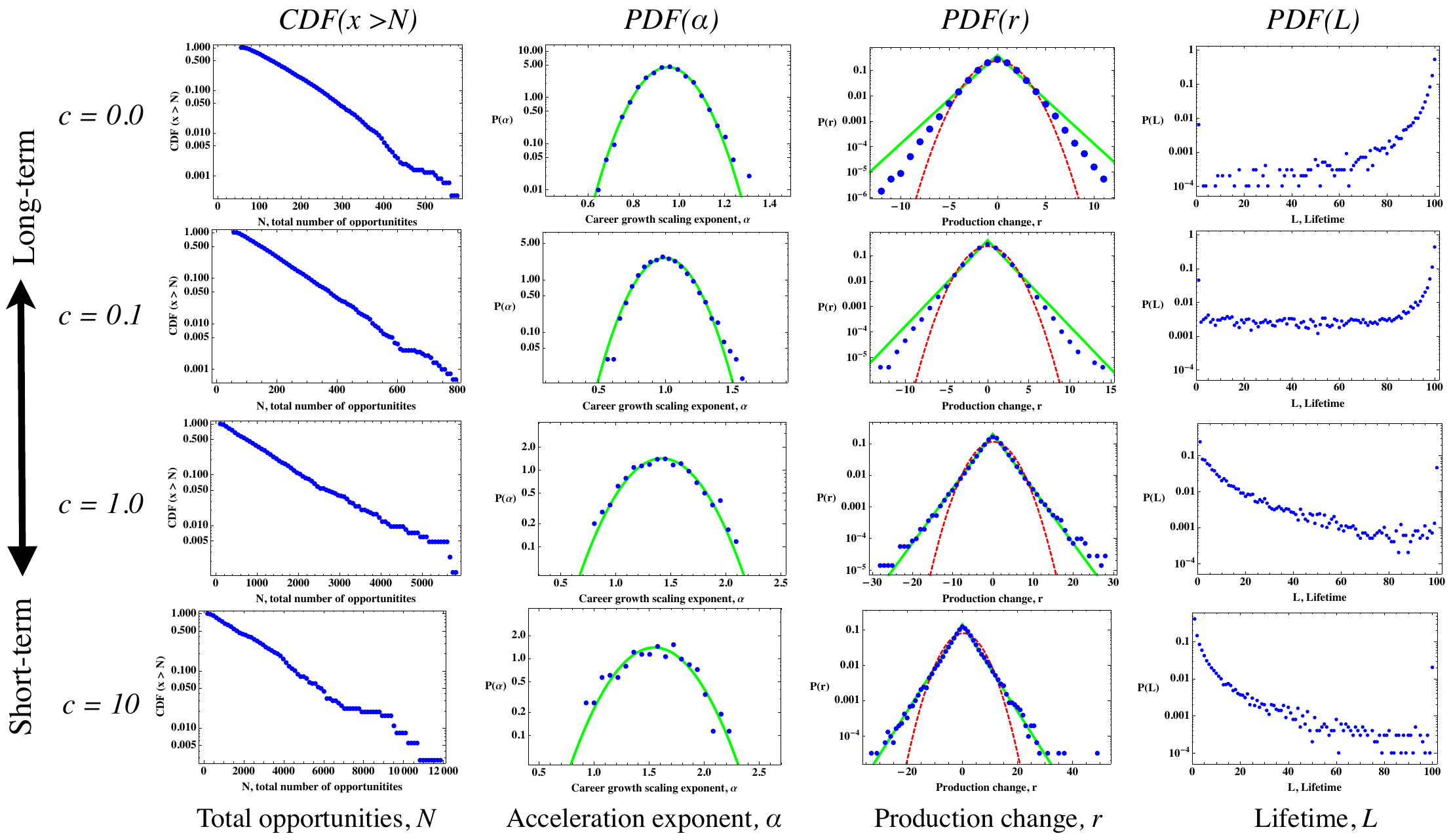}}
\caption{ Monte Carlo simulation of the linear preferential capture model ($\pi = 1$) for varying contract length parametrized by $c$. 
 We plot the probability distributions for (i) $N_{i}$, the total number of opportunities captured by the end period $T$, (ii) the 
 growth acceleration exponent $\alpha_{i}$, (iii) the single period growth fluctuation $r_{i}(t)$ including for comparison the Laplace (solid green) and Gaussian (dashed red) best-fit distributions calculated using the respective MLE estimator, and (iv) the career longevity $L_{i}$ defined as the time difference between an agent's first and last captured opportunity. Results for $c \rightarrow 0$ systems shows that for a ``long-term appraisal'' scenario careers are less vulnerable to low-production
 phases, and as a result, most agents sustain production throughout the career. Conversely,  results for $c\geq 1$ systems show that for a  ``short-term appraisal'' scenario the labor system is driven by fluctuations that can cause career
``sudden death'' for a large fraction of the population. In this short-term appraisal model, there are typically a small number  of agents who are able to capture the majority of  the production opportunities  with remarkably accelerating career  growth reflected by significantly large $\alpha_{i} \geq 1$. Thus,  a few ``lucky'' agents are able to survive the
initial fluctuations and end up dominating the system. In the SI text and Figs. S12-S16, we further show that systems with increased levels of competition ($\pi>1$) mimic systems with short term contracts, resulting 
in productivity ``death traps'' whereby most careers stagnate and terminate early.
 \label{PCmodel}}
\end{figure*}

We use Monte Carlo (MC) simulation to analyze  this 2-parameter model over the course of $t=1...T$ sequential periods. 
 In each production period (i.e. representing a characteristic time to publication), a fixed number
of $P$ production units are captured by the competing agents. 
At the end of each period, we update each $w_{i}(t)$ and then proceed to simulate the next preferential capture period $t+1$.
Since $\mathcal{P}_{i}(t)$ depends on the relative achievements of every agent, the relative competitive advantage of one individual over another is determined by the parameter $\pi$. In the SI Appendix text we elaborate in more detail the results of our simulation of synthetic careers dynamics. We vary $\pi$ and $c$  for a labor force of size $I\equiv1000$ and maximum lifetime $T\equiv 100$ periods as a representative size and duration of a real labor cohort. Our results are general, and for sufficiently large system size, the qualitative features of the results do not depend significantly on the choice of $I$ or $T$.

The case with $\pi = 0$ corresponds to a random capture model that has (i) no appraisal  and (ii) no preferential capture. Hence, in this null model, opportunities are captured at a Poisson rate $\lambda_{p} = 1$ per period. The results of this model (see Fig. S13) shows that almost all careers obtain the maximum career length $T$ with a typical career trajectory exponent $\langle \alpha_{i} \rangle \approx 1$. Comparing to simulations with $\pi >0$ and $c\geq0$, the null  model is similar to a ``long-term'' appraisal system  ($c\rightarrow 0$) with sub-linear preferential capture ($\pi <1$). In such systems,  the long-term appraisal timescale averages out fluctuations, and so careers are significantly less vulnerable to periods of low production and hence  more sustainable since they are not determined primarily by early career fluctuations. 

However, as $\pi$ increases,  the strength of competitive advantage in the system increases, and so some careers are ``squeezed out'' by the larger more dominant careers.
This effect is compounded by short-term appraisal corresponding to $c\approx 1$. In such systems with super-linear
capture rates and/or relatively large $c$,  most individuals experience  ``sudden death'' termination relatively early in the career.
Meanwhile, a small number of ``stars'' survive the initial selection process, which is governed primarily by random
chance, and dominate the system.

We found drastically different lifetime distributions when we varied the appraisal (contract) length
(see Figs.  S12 -- S16). In the case of linear preferential capture with a long-term appraisal
system $c=0$, we find that $10$\% of the labor population terminates before reaching career age $0.94T$ (where $T$ is the
maximum career length or ``retirement age''), and only $25$\% of the labor population terminates before reaching  career age  $0.98T$. On the
contrary, in a short-term appraisal system with $c=1$, we find that $10$\% of the labor population terminates before reaching age
$0.01T$, and $25$\% of the labor population dies before reaching age $0.02T$  (see Table S1).  Hence, in model short contract  systems, the
longevity, output, and impact of careers are largely determined by fluctuations and not by persistence.

Fig. \ref{PCmodel} shows the MC results for $\pi=1$. For $c\geq1$ we observe a drastic shift in the career longevity distribution $P(L)$, which becomes heavily right-skewed with most careers terminating extremely early. This observation is consistent with the predictions of an analytically solvable Matthew effect model \cite{BB2} which demonstrates that many careers have difficulty making forward progress due to the relative disadvantage associated with early career inexperience.
However, due to the nature of 
zero-sum competition, there are a few ``big winners'' who survive for the entire duration $T$ and who acquire a majority of the opportunities allocated during the evolution of the system. Quantitatively, the distribution $P(N)$ becomes extremely heavy-tailed due to agents with $\alpha >2$ corresponding to extreme accelerating career growth. Despite the fact that all the agents are endowed initially with the same production potential, some agents  emerge as superstars following stochastic fluctuations at relatively early  stages of the career, thus reaping the full benefits of cumulative advantage.

\section{Discussion}
An ongoing debate involving academics, university administration, and educational policy makers  concerns the definition
of professorship and the case for lifetime tenure,
 as changes in the economics of  university growth have now placed   {\it tenure}  under the review process
\cite{TenureBook,Tenure}. Critics of tenure argue that tenure  places too much financial risk burden on the modern
competitive research university and diminishes the ability to adapt  to shifting economic,  employment, and scientific
markets. To address these changes,  universities and other research institutes have shifted away from tenure at all levels of
academia in the last thirty years  towards 
meeting staff needs with short-term and non-tenure track positions \cite{TenureBook}. 

For knowledge intensive domains, production is  characterized by {\it long-term} spillovers both through time and through the knowledge network of associated ideas and agents. 
A potential drawback of professions designed around short-term contracts is that there  is an implicit  expectation of sustained annual
production that effectively discounts the cumulative achievements of the individual.
Consequently,  there is a possibility that short-term contracts may reduce the incentives for a young scientist to invest in human and social capital accumulation. 
Moreover, we highlight the importance of an employment relationship that is able to combine positive competitive pressure 
 with adequate safeguards to protect against career hazards and endogenous production uncertainty 
 an individual is likely to encounter in his/her career.
 % Increases in the  duration and improvements to the appraisal
%mechanism for young
%investigator grants can better protect and promote the careers of the diligent and resilient academics towards
%sustainable career growth. 
%An institutional setting that neglects the specific features of academic career trajectories
%may inadvertently  expose temporarily ``cold'' careers, leaving them out to freeze.

In an attempt to render a more objective review process for tenure and other lifetime achievement awards, quantitative
measures  for scientific publication impact  are increasing in use and variety
\cite{AgeDynamicsNobel,gappaper,SciCreditDiffusion,Scientists,GrowthDynamicsH,citationboosts,UnivCite,GrowthScientificOutput}. However, many quantifiable
benchmarks  such as the $h$-index \cite{gappaper}  do not take into account collaboration size or discipline specific
factors. 
Measures for the comparison of scientific achievement should at least account for variable collaboration, publication,
and citation factors \cite{Scientists,UnivCite,GrowthScientificOutput}. Hence, such open problems call for further research into the 
quantitative aspects of scientific output using comprehensive longitudinal data for not just the extremely prolific scientists, but the entire
labor force.

Current scientific trends indicate that there will be further increases in typical team sizes that will forward the emergent complexity arising from group dynamics \cite{TeamScience,MultilevelScience,TeamAssembly}. There is an increasing need for individual/group production measures, such as the output measure
 $Q$, following from Eq. [\ref{gammaefficiency}], which accounts for group efficiency factors.  Normalized production measures which account for coauthorship factors have been proposed  in \cite{Scientists,UnivCite}, but the measures proposed therein do not account for the variations in  team productivity. 
 
The complexity of large collaborations raises open questions concerning scientific productivity and the organization of teams.
We measure a decreasing marginal returns $\gamma <1$ with increasing group size  which identifies  the importance of team management.  
%Instead of relying purely on publication measures, it is important that the review process incorporate scientific contributions in various domains such as teaching, public service, coordination, and
%administration, in addition to research output.
A theory of  labor productivity can help improve our understanding of institutional growth, for organizations ranging in size from  scientific collaborations to universities, firms, and countries
\cite{Growth12,Growth6,MixingGaussians,GrowthScientificOutput,Growth1,Growth11,Growth13}.

\section{Acknowledgements} We thank D. Helbing, N. Dimitri, and O. Penner and an anonymous PNAS Board Member for insightful comments. We gratefully acknowledge support from the IMT and Keck Foundations, the U.S. Defense Threat Reduction Agency (DTRA), Office of Naval Research (ONR), and the NSF Chemistry Division (grants CHE 0911389 and CHE 0908218).

\clearpage
\newpage

\begin{widetext}

\begin{center}
{\large \bf Supporting Information Appendix}
%{\large \bf Supplementary Information}
\end{center}

\bigskip
\begin{center}
{\large \bf Persistence and Uncertainty in the Academic Career} \\
\bigskip
Alexander M. Petersen,$^{1}$ Massimo Riccaboni,$^{2}$, H. Eugene Stanley$^{3}$, Fabio Pammolli $^{1,2,3}$\\
\bigskip
$^{1}$Laboratory for the Analysis of Complex Economic Systems, IMT Lucca Institute for Advanced Studies, Lucca 55100, Italy \\
$^{2}$Laboratory of Innovation Management and Economics, IMT Lucca Institute for Advanced Studies, Lucca 55100, Italy \\
$^{3}$Center for Polymer Studies and Department of Physics, Boston University, Boston, Massachusetts 02215, USA\\
\bigskip
(2012)
\end{center}
\bigskip
\renewcommand{\theequation}{S\arabic{equation}
}
\renewcommand{\thefigure}{S\arabic{figure}}
\renewcommand{\thetable}{S\arabic{table}}
 
\setcounter{equation}{0}  % reset counter 
\setcounter{figure}{0}
\setcounter{table}{0}
\setcounter{section}{0}
\setcounter{page}{1}

\footnotetext[1]{ Corresponding author: Alexander M. Petersen \\
{\it E-mail}: \text{petersen.xander@gmail.com}
}

\section{Data}   
To test the intriguing possibility that competition leads to common growth patterns in complex systems of arbitrary size
$S$, we analyze the production dynamics of two  professions that 
are dissimilar in many regards, but share the common underlying driving force of competition for limited resources. In
order to establish empirical facts that we believe are independent of the details of a given competitive profession, 
we analyze a large dataset of production $n_{i}(t)$ values and corresponding growth fluctuation $r_{i}(t) \equiv n_{i}(t)-n_{i}(t-1)$ values. 
We define the appropriate measures for $n_{i}(t)$ to be (a) the annual number of papers published   by
scientist $i$ and   (b) the seasonal performance metrics of  professional athlete $i$. 
While these two professions both display a high level of competition, they differ in their employment
term structure and salary scale.  In the case of academia, the  tenure system rewards high performance levels with
 lifelong employment (tenure). In contrast,  professional sports are characterized by relatively short contracts that
emphasize continued performance over a shorter time frame and thereby exploit the high levels of athletic prowess in a player's peak years.
The large number of careers in these two professions readily lend themselves to quantitative analysis because the data that quantify the career production trajectory are precisely defined and comprehensive throughout an individual's entire career. Furthermore, because of the
generic nature of competition, we use these two distinct professions to compare and contrast the distribution of  career
impact measures across a cohort of competitors. The datasets we analyze are:

\begin{enumerate}[I :]
\item {\it Academia}:

We analyze the publication careers of 300 physicists which we categorize in 3 subsets each consisting of 100
individuals:
 
\subitem (A) Dataset A corresponds to the 100 most-cited  physicists according to the 
citation shares metric \cite{Scientists} (with average $h$-index $\langle h \rangle = 61 \pm 21$). These 100 careers
constitute 3,951   $r_{i}(t)$ values.

\subitem (B) Dataset B corresponds to the 100 other ``control" scientists, taken approximately randomly from the same
physics  database (with average $h$-index $\langle h \rangle = 44 \pm 15$).  In the selection process for dataset B, we
only consider scientists who have published between 10 and 50  articles in PRL over the 50-year period 1958-2008. These
100 careers constitute 3,534  $r_{i}(t)$ values.

\subitem (C) Dataset C corresponds to 100 Assistant Professors  (with average $h$-index $\langle h \rangle = 15 \pm 7$),
where we select two physicists from each of the top-50 U.S Physics \& Astronomy Departments (according to the U.S. News
rankings). These Asst. Profs. are assumed to be early in their career and relatively accomplished given the difficulty
in obtaining such a position in any given university.  
These 100 careers constitute 1,050  $r_{i}(t)$ values.
\end{enumerate}

In order to control for discipline-specific citation patterns, we select individuals in dataset A and B from  set of all
scientists who have published in {\it Physical Review Letters} (PRL) over the 50-year period 1958--2008. As a measure of
output, we define $n_{i}(t)$ as the number of papers published  
in year $t$ of the career of individual $i$, where year $t=1$ corresponds to the  year of the first publication on
record for author $i$.  
We downloaded the complete publication records of the scientists in datasets A and B from  ISI Web of Science  ({\tt http://www.isiknowledge.com/}) in Jan. 2010, and we downloaded the complete publication records of the scientists in dataset C from  ISI
Web of Science in Oct. 2010. We used the ``Distinct Author Sets''  function provided by ISI in order to increase the
likelihood that only papers published by each given author are analyzed.  \\

\begin{enumerate}
\item[II :] {\it Major League Baseball (MLB):} \\

We analyze 17,292 baseball players over the 90-year period 1920-2009 using comprehensive league data obtained from {\it Sean Lahman's Baseball Archive} accessed at
{\tt http://baseball1.com/index.php}. We  separate the career data into two distinct subsets:
non-pitchers (players not on record as having pitched during a game) and pitchers. 
\subitem (A) For non-pitchers, we analyze two batting metrics: an ``opportunity metric'' - at-bats (AB), and a
``success'' metric -  hits (H). Together, these 8,993 careers constitute 43,043 $r_{i}(t)$ values. 
\subitem (B) For pitchers, we analyze two pitching metrics:  an ``opportunity metric'' - 
innings-pitched measured in outs (IPO), and a ``success'' metric -  strikeouts (K). 
Together, these 8,299 careers constitute 33,965 $r_{i}(t)$ values.

\item[III :] {\it National Basketball Association (NBA):}\\ 

We analyze 3,864 basketball careers, constituting 15,316 $r_{i}(t)$ values, over the 63-year period 1946--2008 using data obtained from  {\it Data Base Sports Basketball Archive} accessed at {\tt http://www.databasebasketball.com/}. 
We analyze two player metrics:
\subitem (A) an
``opportunity metric'' - minutes played (Min.), and 
\subitem (B) a ``success'' metric - points scored (Pts.) 
\end{enumerate}  

Since sports careers  typically peak for athletes around age 30, we account for a time-dependent career trajectory which
is dominant in most sports careers by   ``detrended" the measures for career growth fluctuations.
In the case where we do not account for a individual fluctuation scale,
\begin{equation}
R_{i} \equiv [r_{i}(t)-\overline{r}(t)] / \sigma(t) \ . 
\end{equation}
In this case we detrend with respect to the average production  difference $\overline{r}(t)$  and the standard deviation  of
production difference $\sigma(t)$ which are calculated using all careers from a given sports league, conditional on the career year $t$.

In the case where we do account for individual variations, we first define
$z_{i}(t) \equiv (r_{i}(t)-\langle r_{i} \rangle)/\sigma_{i}$ to be normalized with respect to
the individual career scales $\langle r_{i} \rangle$ and $\sigma_{i}$ which are the average and standard deviation of the
production change of athlete career $i$. Then we define the detrended growth rate as
\begin{equation}
R'_{i} \equiv  [z_{i}(t)- \langle z(t) \rangle ] / \sigma_{z(t)} \ ,
\end{equation}
where in this case we detrend with respect to the average $\langle z(t) \rangle$ and standard deviation $\sigma_{z(t)}$  calculated by 
collecting all $z_{i}(t)$  values for a given career year $t$. This detrending better accounts for the relatively strong time-dependent growth patterns in sports.

In this section we analyze the annual production of scientists measured as  the number of papers published $n_{i}(t)$ over the period of a year. 
Using this measure does not account for the variability in the length of production, say in the number of pages, nor does it account for the impact of the paper,
a quantity  commonly approximated by a paper's citation number. Instead, we consider a 
simple definition that a scientific product is a final output of a collection of inputs. Furthermore, in science
 it is assumed that the peer review process establishes a quality threshold so that only manuscripts above a certain quality and novelty 
 standard can be published and incorporated into the scientific body of knowledge.
 
 Prior theories of scientific production have also used the number of publications as a proxy for scientific output. In particular,
 the Shockely model \cite{ShockleyProductivity} 
 proposed a simple multiplicative factor model for the production $n_{i}(t)$
 which predicts a log-normal distribution for $P(n)$. An alternative null model   for $n_{i}(t)$ is the Poisson process, which assumes that each 
 individual is endowed with a rate parameter $\omega$ related to an individual's production factors. This model predicts a Poisson distribution for $P(n)$.
 However, a shortfall of these models is that multiplicative parameters in the Shockley model and the rate parameter $\omega$ are difficult to measure, especially 
 if the set of individuals span a large range of production factors, and moreover,  if the careers are non-stationary.
 
 Fig. \ref{nPDF} shows the unconditional probability distribution $P(n)$ calculated by aggregating all $n_{i}(t)$ values for all scientists and all years into an
 aggregate dataset. Naively, the distributions are well-fit by the Log-normal distribution, and so there is an apparent agreement with the multiplicative factor Shockley model.
However, the distribution $P(n) = \sum_{i=1}^{100} P(n|S_{i})$ is the aggregate distribution constructed from 100 individual career trajectories $n_{i}(t)$, each with 
varying size $S_{i}$. Indeed, we demonstrate in 
Figs. \ref{NTi} and \ref{NTishocks}
 to be non-linear, with time-dependent residuals around the moving average. Hence, it is not possible from the unconditional pdf $P(n)$ to determine if the 
 process underlying scientific production corresponds to a simple multiplicative process or a Poisson process. 
 
 In order to better account for the variable size $S_{i}$ of each career which affects the rate at which an individual is able to capture publication opportunities,
 we plot in Fig. \ref{ScalednPDF} the pdf of the normalized output
 \begin{equation}
 Q_{i}=\frac{n_{i}(t)}{f_{i}(k)} \ .
 \label{Qdef}
 \end{equation}
We calculate the normalization factor $f_{i}(k) = q_{i}[k_{i}(t)]^{\gamma_{i}}$ for each individual $i$ by estimating the parameters $q_{i}$ and $\gamma_{i}$ for 
each scientist $i$ from the single-factor model
 \begin{equation}
 n_{i} = q_{i}k_{i}^{\gamma_{i}} \ .
 \label{fdef}
 \end{equation}
 where $n_{i}(t)$ is the annual production in year $t$ and $k_{i}(t)$ is the total number of distinct coauthors in year $t$.
 Hence, $Q_{i}$ represents the production factor above $Q>1$ or below $Q<1$ what would be expected from the author $i$ given the fact that he/she had additional
 inputs from $k_{i}(t)-1$ individuals that year. 
 This model assumes that the major component contributing to production is the collaboration degree $k$ of the research output, and also 
 assumes that the input of each coauthor contributes equally to the final output. Clearly, these assumptions neglect some important idiosyncratic details
 affecting scientific publication, but given the incomplete information associated with every publication, it is a
 decent approximation.  We estimate $q_{i}$ and $\gamma_{i}$ 
 by performing a linear regression of $\log n_{i}$ and $\log k_{i}$ using the first $L_{i}$ years of each career, neglecting years with $n_{i}=0$. We use
 $L_{i}=35$ years for dataset [A] and [B] scientists, and $L_{i} = 10$ years for dataset [C] scientists.
 
 In Fig. \ref{NSDscaling}(c) we approximate $\gamma$  using all $n(t)$ within each dataset with $k\leq 50$, and performing a regression of the model 
  \begin{equation}
\ln  n  = \ln q + \gamma \ln k + \epsilon 
 \label{gammacalc}
 \end{equation}
to estimate $\gamma$,  where $\epsilon$ is the residual due to other unaccounted production factors.
For each dataset we find that the aggregate efficiency parameter $\gamma$ is approximately equal 
to the average $ \langle \gamma_{i}\rangle$ calculated from  the 100 $\gamma_{i}$ values in each career dataset:
 $\gamma = 0.68 \pm 0.01$ [A], $\gamma = 0.52 \pm 0.01$ [B], and $\gamma = 0.51 \pm 0.02$ [C].
 Furthermore, the  $\psi \approx  \gamma$ since the size-variance scaling parameter $\psi$ is also an efficiency 
 measure that relates the scaling of output $n$ to input $k$.
 
 As a result of this analysis, we quantify the scaling exponent $\gamma<1$ of the decreasing 
 marginal returns in the scientific production function for projects with $k\leq 50$. This likely stems from the inefficient management costs associated with large group collaborations which
 typically manifest in a larger production timescale.  In fact, for years with $k\geq 50$ coauthors, scientific output shows decreasing returns to scale.
 Interestingly, the star scientists in dataset [A] display significantly larger efficiency, quantitatively showing the importance of  management skills in scientific success.
 
 The normalized production values are normalized to  units of ``expected production" conditional on the $k_{i}$ inputs for author $i$. We 
 aggregate all data from each dataset and 
 show in Fig. \ref{ScalednPDF}   that the $Q$ values are well-described by the  Gamma distribution 
 \begin{equation}
 P(Q) =  Q^{m-1} \frac{\exp[-Q/\theta]}{\theta^{m} \Gamma(m)}
 \end{equation}
 where $m$ is the shape parameter and $\theta$ is the scale parameter.
 Surprisingly, we find that dataset [A] and [B] have approximately equal Gamma parameters, 
 indicating that besides their production efficiency, top scientists are virtually indistinguishable with average normalized output  $\langle Q \rangle = m \theta >1$.
 For each dataset we calculate the Gamma parameters using the maximum likelihood estimator method: $m=5.45$ and $\theta = 0.21$ [A], $m=5.60$ and $\theta = 
 0.20$ [B], and $m =7.00$ and $\theta = 0.15$ [C].
  We leave it as an open question to determine  
 why the Gamma distribution describes so well the production statistics. 
 We ponder the intriguing possibility that the 
 stochastic dynamics underlying individual production corresponds to an increasing L\'evy process with variable jump length  which is 
 known to produce a Gamma distribution.  
 
 %Surprisingly, we find that the distribution of normalized production $Q$ shown in Fig. \ref{ScalednPDF} is not distributed as a Poisson or Log-normal distribution, as in alternative models for production, but as a Gamma distribution. 
% This leads to the intriguing possibility that a  Gamma L\'evy process may provide a theoretical framework for the bursts of productivity that underly the production of creative knowledge.
 
 \clearpage
 \newpage
 
 \section{Quantifying the Career Trajectory} 
 The reputation of an individual is typically cumulative, based on the total
sum of achievements, which we approximate by the  cumulative output $N_{i}(t)$  (e.i. number  of papers published by
year $t$).  
%One possible measure for career achievement is the cumulative production $N_{i}(t)$ of an individual $i$ in year $t$ of
%their career. 
In Figs. \ref{NTi} and \ref{NTishocks} we plot $N_{i}(t)$ for several individuals. The careers presented in Fig.
\ref{NTi} are more linear, 
indicating quantifiable career trajectory that has the approximate form
\begin{equation}
N_{i}(t) = \sum_{t'=1}^{t} n_{i}(t')  \approx A_{i} \ t^{\alpha_{i}}   \  ,  \ \ t < T_{i} 
\label{powlawN}
\end{equation}
where $n_{i}(t)$ are the number of papers in year $t$ of the scientist's career which begins with $t \equiv 1$ in the
year of his/her first publication, and begins to decline around time $T_{i}$ which is the time horizon over which the
scaling regularity  holds before termination and aging effects begin to dominate the career. 
In our analysis of academic career trajectories $N_{i}(t)$, we only analyze $N_{i}(t)$ for $t\leq40$ years in order to
account for such termination affects.

The smooth career trajectories which appear as a linear curve when plotted on log-log scale are characterized by  an
amplitude parameter $A_{i}$ and a scaling exponent $\alpha_{i}$. However, as indicated by Fig. \ref{NTishocks}, there
are also non-stationary $N_{i}(t)$ which are dominated by ``career shocks'' that significantly alter the career
trajectory. Such career shocks have been demonstrated using publication impact measures (e.i. citations, and h-index
sequences) \cite{GrowthDynamicsH, citationboosts, AgeDynamicsNobel}, and here we show that they even occur at the more fundamental level
of individual production dynamics.

In order to analyze the characteristic properties of $N_{i}(t)$ for all 300 scientists analyzed, we define the normalized
trajectory $N'_{i}(t) \equiv N_{i}(t) / \langle n_{i} \rangle$, where $\langle n_{i}(t) \rangle$ is the average annual
production rate of author $i$, and so by construction $N'_{i}(L_{i}) = L_{i}$. 
Fig. \ref{NT}(A) shows the characteristic production trajectory obtained by averaging the 100 individual $N'_{i}(t)$ for
each dataset,
\begin{equation}
 \langle N'(t) \rangle \equiv \Big \langle \frac{N_{i}(t)}{ \langle n_{i} \rangle} \Big \rangle \equiv \frac{1}{100}
\sum_{i=1}^{100} \frac{N_{i}(t)}{ \langle n_{i} \rangle}  \ .
\end{equation}
The standard deviation $\sigma( N'(t))$ is shown in Fig. \ref{NT}(B), which has a broad peak that is a likely signature
of career shocks that can significantly alter the career trajectory. The characteristic trajectory for each dataset are
well-approximated by the scaling relation 
\begin{equation}
\langle N'(t) \rangle \sim t^{\overline \alpha}
\end{equation}
with characteristic scaling exponents $ \overline \alpha>1$ that are significantly greater than unity:
 $ \overline \alpha= 1.28\pm0.01$ for Dataset A, $ \overline \alpha =1.31\pm0.01$ for Dataset B, and $ \overline
\alpha=1.15\pm0.02$ for Dataset C. 
 This fact implies that there is a significant cumulate advantage in scientific careers which allows for the career
trajectory to be accelerating.
 In Fig. \ref{NT}(C) and \ref{NT}(D) we plot the analogous $\langle N'(t) \rangle$ curves for professional sports
metrics, where for this profession, $\overline \alpha \approx 1$ for 
 all measures analyzed. This quantitative feature is likely due to the fact that annual production in professional sports is capped by the
limited number of opportunities provided by a season, whereas in academics, the number of publications a scientist can
publish is in principle unlimited. Also, in more labour-intensive activities are likely to experience smaller returns   since physical labor is non-cumulative with  less
spillover through time.
 
In Fig. \ref{NtotalScaled} we plot each
individual career trajectory using the rescaled time $t'_{i} = t^{\alpha_{i}}$ as an additional visual test of the scaling model given by Eq. \ref{powlawN}.  We show that on average, all curves
$i=1..300$ approximately collapse onto the expected curve $N_{i}(t)/A_{i} = t'$, where the residual difference 
$ \epsilon_{i}(t') \equiv N_{i}(t)/A_{i}-t'$ are  likely due to career shocks of various magnitudes.  We plot the
average and standard deviation of each set of 100  $N_{i}(t)/A_{i}$ curves which show that most of the shocks
$\epsilon_{i}(t')$, with some significant exceptions, lie within the 1$\sigma$ standard deviation denoted by the error
bars. In Fig. \ref{Alpha} we plot the probability distributions $P(\alpha_{i})$ for each academic dataset. For each
dataset, the average value $\langle \alpha_{i} \rangle$ is in good agreement with $\overline \alpha$, the scaling
parameter calculated for the corresponding trajectory $\langle N'(t) \rangle$.  

 \clearpage
 \newpage
 
 \section{Exponential Mixing of Gaussians}

 The idea that entities are independent and identically distributed is an unrealistic assumption commonly made in analyses of complex systems. 
 The unconditional pdf  $P(r)$ is commonly analyzed in
 empirical studies where insufficient data are  present   to define normalized 
$r_{i}'$ measures for each sample constituent $i$.
 Nevertheless, when modeling the evolution of complex based on empirical data corresponding to distinct subunits (such as individual careers, companies, or nation regions),
  unconditional quantities that account for variations in underlying
production factors should be used.

 In the case of scientific output, there are many 
production factors that combine together and determine the amount of human efforts needed to produce a unit of production. 
In general, consider the value $f_{i,j}$ of individual $i$  corresponding to  his/her relative abilities in the production factor $j=1...J$ corresponding to a variety of attributes: knowledge, genius,
persistence, reputation, mental and physical health, communication skills, organization skills, and access to technology, equipment and data, etc.
In this study, we compare
scientists who publish in similar journals. Still,  the scientific input required for each scientific output can vary by
a
large amount, largely depending on the technology needed to perform the analysis, ranging from particle accelerators to  just a pencil and paper.

In a very generalized representation, an unconditional distributions $P(r)$, such as  shown in Fig. \ref{dNPDF}(a-d) for production change $r$, may 
follow from a mixture of conditional  Gaussian distributions $P(r|S_{i})$
\begin{equation}
P_{\psi}(r) = \int_{0}^{\infty}  P(r|S) P(S) dS  \approx \sum_{i=1}^{I}P_{i}(r|S_{i}) P(S_{i}) \ .
\label{expmix0}
\end{equation}
The underlying conditional distributions are characterized by the average $\langle r \rangle_{S_{i}}$ and variance $\sigma_{i}^{2} \equiv VS_{i}^{\psi}$  
\begin{equation} 
%P(r| S) = \exp[ -r^{2}/2\sigma^{2}(r)] / \sqrt{2\pi \sigma^{2}(r)} \ .
P(r| S_{i}) = \exp[ -(r-\langle r \rangle)^{2}/2VS_{i}^{\psi}] / \sqrt{2\pi VS_{i}^{\psi}} \ .
\end{equation}
which are each parameterized by the characteristic collaboration size $S_{i}$. 
In cases where the average change $\langle r \rangle \approx 0$,  then the  distribution $P(r| S_{i})$ is characterized by only the fluctuation scale $\sigma_{i}(r)$.
Fig. \ref{drnormPDF} demonstrates that the 
normalized production change $r'_{i}(t) = (r-\langle r_{i} \rangle)/\sigma_{i}$  is distributed according to a Gaussian distribution.
Hence, using normalized variables, we have mapped the process to a  universal scaling distribution $P(r|S_{i})$.

When the distribution $P(S_{i})$  is exponential, 
\begin{equation}
P(S_{i}) = \lambda e^{-\lambda S_{i}}
\end{equation}
then mixture is termed an ``exponential mixture of Gaussians''  \cite{MixingGaussians}, where the units have
 characteristic size $\overline{S_{i}} = 1/\lambda$. 
 Fig. \ref{NAvePapersCDF} shows that the distribution of 
collaboration radius $S_{i}$ is approximately exponential for each dataset, supporting the case for exponential mixing.
Using the cumulative distribution of $S$ for each data set   we calculate  $\lambda = 0.15 \pm 0.01$ [A], 
$\lambda = 0.11\pm 0.01$ [B], and $\lambda = 0.11 \pm 0.01$ [C]. 
While the tail behavior of $P(r)$ can be used to better discriminate the value of $\psi$, we do not have sufficient data
in this analysis to perform a more rigorous test of the tail dependencies, or in general, to investigate the
distribution of significantly  large $r_{i}(t)$ values.

The scaling relation $\sigma_{i}(r) \sim S_{i}^{\psi/2}$   determines the functional form of the 
aggregate $P_{\psi}(r)$. 
Clearly, $\sigma(r)$ increases for $\psi>0$ values,  whereas for values $\psi<0$, $\sigma(r)$ decreases with size $S_{i}$. This latter case is empirically observed for countries and firms  \cite{Growth11}, whereby in general,  large economic entities are able to decrease growth volatility by increasing and diversifying their portfolio of growth products.
In our analysis of scientific careers  we define $S_{i} \equiv Med[k_{i}(t)]$, the median number of
distinct coauthors per year, as a proxy for the ability of the career to attract new opportunities,  and hence, as a proxy for the size $S_{i}$ of an academic career. For professional athletes, we define the career size as the average number of points scored over the career $S_{i} \equiv \langle p_{i}(t) \rangle$. 
In Fig. \ref{NSDscaling} we calculate  $\psi/2 \approx 0.40\pm 0.03$ (regression coefficient $R =0.77$) for dataset [A], $\psi/2 \approx 0.22 \pm 0.04$ ($R = 0.51$) [B], and $\psi/2 \approx 0.26 \pm 0.05$ ($R =0.45$) [C]. 

The role of mental, physical, and group spillovers is quite different in professional sports. Athletes attract future opportunities largely through their 
historical track record, which is heavily
weighted on performance in the near past, and less on the cumulative history. Hence, for this performance-based labor
force, we use a simple definition of  ``team value''   to define the career size $S_{i}$. This quantity is easier to
define for basketball, since there are smaller  differences between players of different team position than in other sports.  For NBA player
$i$ we define $S_{i}$ as   the average number of points scored per year,  $S_{i} \equiv \langle p_{i} \rangle$.  
Fig. S9 shows a crossover value $S_{c}$ which we interpret to reflect the fact that sports players typically fall into one of two categories: starters (everyday players) and replacement (game filler) players. We calculate  $\psi/2 \approx 0.38 \pm 0.02$ for emerging and ``second string'' careers with  
$S_{i}< S_{c}$, and a decreasing size variance relation  ($\psi<0$) for high-value careers with $S_{i}>S_{c}$.  Similar
values occur in the MLB. These two  $\psi$ regimes reflect the crucial balance of risk and reward in  short-term contract professions.

A variety of pdfs $P_{\psi}(r)$ can result from the  exponential mixture of
Gaussians
\begin{equation}
P_{\psi}(r) = \int_{0}^{\infty} \lambda e^{-\lambda S} \frac{1}{\sqrt{2\pi \sigma^{2}(r)}} \exp[ -r^{2}/2\sigma^{2}(r)] dS
\label{expmix}
\end{equation}
depending on the value of $\psi$ which quantifies the size-variance relation. 
The functional form of $P_{\psi}(r)$ can vary in both the bulk and the tails of the distribution \cite{MixingGaussians}.
A simple result  which follows from the case $\psi =1$ is  the Laplace  (double-exponential) distribution
\begin{equation}
P_{\psi = 1}(r) = \sqrt{\frac{\lambda}{2V}} \exp\Big[-\sqrt{\frac{2 \lambda}{ V}} \vert r  \vert \Big] \ .
\label{LaplaceEqn}
\end{equation}
This distribution is a member of the  family of Exponential power distributions which follow from the range of values $\psi \geq 0$ \cite{MixingGaussians}.
In general, if the scaling values are in the range
$\psi \geq 0$, then the exponential mixture leads to an Exponential power distribution 
\begin{equation}
P(r) = \frac{\beta}{\sqrt{2} \sigma \Gamma (1/ \beta)} \exp[ -\sqrt{2}(\vert r   \vert /
\sigma)^{\beta}]
\end{equation}
with shape parameter $\beta$ in the range $\beta  \in (0,2]$ \cite{MixingGaussians}. The pure exponential $P(r)$
with  $\beta =1$ corresponds to the case $\psi = 1$. 
The pure Gaussian $P(r)$
with  $\beta =2$ corresponds to the case $\psi = 0$. 

 %In Figs.~\ref{dNPDF} (a-d) we show the empirical probability density function (pdf) computed from the $r_{i}$ values of
%scientists and the $R_{i}$ values of athletes. 
%For all career data analyzed here,  we find a remarkable regularity characterized by a ``tent-shaped''  $P(r)$ when we
%plot the distribution on semi-logarithmic axes. This result follows from the fact that in these professions $\psi \approx 1$ and $P(S)\approx \lambda \exp[ -\lambda S]$.

Furthermore, if the annual production is logarithmically related to an underlying
production
potential, $n_{i}(t) \propto \ln U_{i}(t)$, then $r_{i}(t)  \propto \ln U_{i}(t) - \ln U_{i}(t-1)$ quantifies the logarithmic change (``growth rate'') of $U_{i}(t)$. This forms
the analogy with growth dynamics of large institutions with  size $S\gg1$. For example, in the case of financial securities such as the
stock of a company $i$,  the growth rate $r_{i}(t)$ measure the logarithmic change in the market's expectations of the
company's future earnings potential captured by the market capitalization and price  \cite{Growth13}. 
As a result, 
distributions $P(r)$ of career growth fluctuation $r$, which we plot in  Figs.~\ref{dNPDF} (a-d),  can be seen as a bridge between the micro level and the macro level of economic growth fluctuation.
  A theory of  micro growth processes can help improve  the growth forecasts for economic organizations ranging in size from  scientific collaborations to universities and firms
\cite{Growth1,Growth6,Growth11,Growth12,Growth13, MixingGaussians,GrowthScientificOutput}.

 \section{Nonlinear preferential capture  model} 

\begin{comment}
TODO:
\begin{enumerate}
\item Figure out some prediction of model that can be tested against data. What is $S_{i}$ in model?
\item In model, justify as preferential attachment w/out random hazards (as in Matthew paper) to show that even w/out random removal, the career length distribution shows the detrimental affects of such a high level of competition.
\item Include panel of figures that show output for varying c value and gamma value: P(r|c,gamma), P(ntotal | c, gamma), P(alpha | c, gamma) just to give a hint of what the model can reproduce.
\end{enumerate}
\end{comment}

Here we describe a stochastic system in which a finite number of opportunities are distributed to a system of individual competing agents $i = 1...I$. The opportunities are distributed  in batches of $P$ opportunities per arbitrary time interval.  This model  has two parameters. \\

(i) $\pi$ determines the preferential capture mechanism (the value $\pi=1$ corresponds to the traditional ``linear'' preferential attachment model) and \\

(ii) $c$ determines the performance timescale $1/c$ which is incorporated into the calculation of the capture rates of each individual. The value $c=0$ corresponds to a long-term memory and $c \gg 1$ corresponds to short-term memory. \\

\noindent We use this simple model to show that a system governed by a preferential capture can become dominated by fluctuations when $c$ is large. The value $1/c$ quantifies the ``performance appraisal timescale'': a small $c$ corresponds to a labor system  with long contracts, 
or some alternative mechanism that provides employment insurance through periods of low production, so that the ability to attract future opportunities is largely based on the cumulative record of career achievement. Conversely, a large $c$ corresponds to a labor system with short contracts   in which the ability to attract future opportunities is largely based on the accomplishments in the near past, requiring an agent to maintain relatively high levels of production in order to survive. In this latter case, we find that (natural) fluctuations in the annual production 
can cause a significant fraction of the careers to ``fizzle out'' leaving  behind only a few ``super careers'' who attract almost all of the opportunities. In other words, short contracts can tip the level of competition into dangerous territory whereby careers are largely determined by fluctuations and not persistence.

%The goal is to construct a simple model for the capture of output opportunities by agents that can explain the empirical $\psi$ value and the relation between the standard deviation $\sigma_{i}(r)$ and the definition of size $S_{i}\equiv \langle n_{i}(t) \rangle$ as in our paper.\\

\subsection{System of competing agents}

\begin{itemize}
\item [1)] The system consists of $I \equiv 1000$ agents competing for $P$ opportunities that are allocated in a single period. There is no entry, hence the number $I$ is kept constant. Also,  $P$ is also kept constant, so there is no growth in the labor supply.
\item [2)] We run the Monte Carlo (MC) simulation for  $T \equiv 100$ time periods  and all agents are  by construction from the same age cohort (born at same time).
%\item [3)] %The initial production potential $n_{0,i}$ of agent $i$ is drawn from an exponential distribution with exponential scale parameter $\lambda = 0.1$ as measured in the paper. Hence, I assume that $\langle n(t) \rangle \approx n_{0,i}$ since $\alpha \approx 1$ (or is not {\it much} larger than 1) and so the assumption is that the average career output is a good proxy for  an individual's initial intrinsic potential. This discounts the possibility of career shocks significantly altering the career, which is certainly a strong factor empirically.
\item [3)] Each time period corresponds to the  allocation of $P \equiv \sum_{i=1}^{I}n_{0,i}$ opportunities, sequentially one at a time, to randomly assigned agents $i$, where $n_{0,i} \equiv 1$ is the potential production capacity of a given individual. 
\item[4)] The assignment of a given opportunity is proportional to the time-dependent weight (capture rate) $w_{i}(t)$ of each agent. Hence, the assignment of 1 opportunity to agent $i$ at period $t$ results in the production (achievement) $n_{i}(t)$ to  increase by one unit: $n_{i}(t) \rightarrow n_{i}(t)+1$. In the next time period $t+1$, we update  the weight $w_{i}(t+1)$ to include the performance $n_{i}(t)$ in the current period.
\end{itemize}

\subsection{Initial Condition}

The initial weight at the beginning of the simulation is $w_{i}(t=0) \equiv n_{c}$ for each agent $i$ with $n_{c}\equiv 1$. The value $n_{c} >0$  ensures that competitors begin with a non-zero production potential, and corresponds to a  homogenous system where all agents begin with the same production capacity. Hence, we do not analyze the more complicated  model wherein external factors (i.e. collaboration factors) can result in a heterogeneous production capacity across scientists. By construction, each agent begins with one unit of achievement $n_{i}(t=1) \equiv 1$.
%I will  further explore the system  assuming homogenous  initial condition  $w_{i}(0) \equiv n_{c}$ for varying $n_{c}$. \\

\subsection{System Dynamics}
\begin{itemize}
 \item [1)]  In each Monte Carlo step we allocate one opportunity to a randomly chosen  individual $i$ so that $n_{i}(t) \rightarrow  n_{i}(t)+1$
 
\item [2)] The individual $i$ is chosen with probability $\mathcal{P}_{i}(t)$ proportional to $[w_{i}(t)]^{\pi}$ 
\begin{equation}
\mathcal{P}_{i}(t) = \frac{w_{i}(t)^{\pi}}{\sum_{i=1}^{I} w_{i}(t)^{\pi}}
\label{Pi}
\end{equation}
where the value $w_{i}(t)$ is given by an exponentially 
weighted sum over the entire achievement history
\begin{equation}
w_{i}(t) \equiv \sum_{\Delta t =1}^{t-1} n_{i}(t-\Delta t) e^{-c\Delta t} \ .
\label{wit}
\end{equation}
%$w_{i}(t) \equiv [cN_{i}(t)+(1-c)n_{i}(t-1)]^{\pi}$ where $N_{i}(t) \equiv \sum_{t'}^{t}n_{i}(t)$
% is the total production up to time $t$ and $n_{i}(t)$ is the net production in the previous period. 
The parameter $c \geq 0$ is a memory parameter which determines how the record of  accomplishments in the past  affect the ability to obtain new opportunities in the current period, and therefore, the future.
The limit $c = 0$ rewards long-term accomplishment by equally weighting the entire history of accomplishments. Conversely, when $c \gg 1$ the value of $w_{i}(t)$ is largely dominated by the performance  $n_{i}(t-1)$ in the previous period,   corresponding to increased emphasis on  short-term accomplishment in the immediate past. Intermediate values  $0  < c < 1$  weight more equally  the immediate past and the entire history of accomplishment.
\item [3)] The exponent $\pi$ determines  how the relative ability to attract opportunities $\mathcal{P}_{i}/\mathcal{P}_{j} = [w_{i}(t)/w_{j}(t)]^{\pi}$ depends on  the weights $w_{i}(t)$ and $w_{j}(t)$  between two individuals $i$ and $j$. The linear capture case follows from $\pi = 1$, uniform capture $\pi = 0$,  super linear capture $\pi >1$, and sub-linear capture $\pi < 1$.
\item [4)]  At the end of each time period, the weight $w_{i}(t)$ is recalculated and used for the entirety of the next MC time period corresponding to the allocation of the next $I \times n_{c}$ achievement opportunities.
\end{itemize}

\subsection{Model Results}

We simulate this system for a realistic labor force size $I=1000$ with the assumption that in any given period, an
individual has the capacity for  one unit of production ($n_{c}\equiv 1$). We evolve the system for $T=100$ periods
corresponding to $I \times n_{c} \times T$ Monte Carlo time steps. The timescale $T$ represents the  (production)
lifetime of  individuals with finite longevity. In this model we do not include exogenous  shocks  (career hazards) that
can result in career death \cite{BB2}. Here we  analyze four quantities:
\begin{itemize}
 \item [1)]  The distribution $P(N)$ of the total number of opportunities  $N_{i}(T)\equiv \sum_{t=1}^{T} n_{i}(t)$
captured by agent $i$ over the course of the $T-$ period simulation.
 \item [2)]  The distribution $P(\alpha)$ of the career trajectory scaling exponent   $\alpha_{i}$ defined in Eq.
\ref{powlawN} which quantifies the (de)acceleration  of production over the course of the career.
  \item [3)]  The distribution $P(r)$ of production outcome change   $r$ defined in Eq. \ref{Dn} which quantifies the
size of endogenous production shocks.
  \item [4)]  The distribution $P(L)$ of career length $L_{i}$ which measures the active production period of each
career starting from $t=0$. We define activity as the  largest period value $L_{i}$ for which $n_{i}(L_{i})=0$,  which
in other words,  corresponds to truncating all $0$ production values from the end of the trajectory $n_{i}(t)$ and
defining $L_{i}$ as the length of this time series. 
  \end{itemize}
  
  We display these four distributions, from left to right, for varying $\pi$ and $c$ values, in each panel of Figs. 
\ref{pi0C0} -- \ref{C10p0}. Empirical distributions calculated from MC simulations are plotted as blue dots, with
benchmark
distributions described below plotted as solid green curves. For each $\pi$ and $c$ value we simulate 10
MC systems, and combine the results into aggregate distributions which are shown. For
simulations with $\pi >1$ the pdf data are aggregated over the results of 50 MC simulations. We
 list below some of our main
 observations.\\

 For $\pi =1$, independent of $c$, we observe exponential $P(N)$, consistent with the prediction of the linear
preferential capture model in the case of no firm entry ($b=0$) in the model of Kazuko et al.~\cite{ExpDistSize}.
However, the distribution $P(L)$ and the distribution $P(\alpha)$ does depend strongly on $c$, reflecting the
possibility of career ``sudden death'' for large $c$.\\

 For the  $P(\alpha)$ distributions (middle-left panels), the solid green line is a best-fit Gaussian distribution
(using the MLE method)
for the set of $\alpha_{i}$ values computed for careers that did not undergo ``sudden death.''  \\

 For the $P(r)$ distributions (middle-right panels), the solid green curve corresponds to a best-fit Laplace
distribution (using the MLE method) and the dashed red curve corresponds to a best-fit Guassian distribution (using
the MLE
 method) which we show only
 for benchmark comparison. Typical empirical distributions (values shown as blue
dots) range from being distributions that are Gaussian to distributions that are Laplacian in the bulk but  with heavy
tails.\\

 For the $P(L)$ distributions (right most panels), we note that the most likely career length $L$ is typically either
$L=1$ or $L=T$ for all systems analyzed. However, there are likely   $c$ and $\pi$ parameter values corresponding to 
$P(L)$ that is uniform distributed over the entire range of $L$ values, which may be an interesting class of system to
analyze in future analyses since such a system promotes diversity across the entire longevity spectrum. The system we
show for $\pi = 1.2$ and $c=1$ appears to be close to this scenario. \\

  Fig. \ref{pi0C0} shows the null model with no preferential capture ($\pi = 0$). We confirm that the careers in this
model are driven by a stochastic accumulation process that is equivalent to a Poisson process with rate $\lambda_{p}\equiv 1$. In this homogenous
system,  each career gains on average one opportunity each time period, so that at the end of the simulation, the
distribution $P(N)$ is a Poisson distribution with $\langle N\rangle = \lambda_{p} T$ (shown as the solid blue line) which
  fits the model data excellently. For these careers, the typical $\alpha = 1$, the production changes are
well-approximated by a Gaussian distribution, and most careers are sustained for the maximum possible lifetime
corresponding to $T$ periods.\\
  
   Fig. \ref{C0}  shows the system with $c=0$ corresponding to comprehensive  career appraisal corresponding to a
long-term memory system. We analyze this system for 4 values of $\pi = {0.8, 1.0, 1.2, 1.4}$. This ``long-term memory''
scenario corresponds to a long-term contract profession whereby careers are less vulnerable to periods of low
production. As a result, most careers sustain production throughout the career.\\
   
Fig. \ref{C0p1}  shows the system with $c=0.1$ corresponding to  an effective memory timescale of $1/c =10$ periods. We
analyze this system for 4 values of $\pi = {0.8, 1.0, 1.2, 1.4}$. This ``medium-term memory'' scenario yields a rich
variety of careers for $\pi =1$, but for $\pi =1.2$ the system becomes quickly dominated by ``rich-get-richer'' effects
which results in careers being vulnerable to low production fluctuations.\\
      
Fig. \ref{C1p0} shows the system with $c=1$ corresponding to  an effective memory timescale of $1/c=1$ period. 
We analyze this system for 4 values of $\pi = {0.8, 0.9, 1.0, 1.1}$. For all values of $\pi$ analyzed, we observe a
system that is dominated by careers that are cut short by the high levels of competition induced by the relatively high
value placed on continued production. \\
  
Fig. \ref{C10p0} shows the extreme case of a ``no memory'' scenario  in which $w_{i}(t) \approx n_{i}(t-1)$ whereby most
careers experience sudden death due to endogenous negative production shocks early in their career. The lucky few
careers who survive this period end up as  rich-get-richer ``superstars.'' This behavior occurs for all systems analyzed
using 4 values of $\pi = {0.8, 0.9, 1.0, 1.05}$.\\

  \subsection{Discussion of the model in relation to the Academic labor market}
  
  One serious drawback of short-term contracts are the tedious  employment searches, which displace career momentum by
taking focus energy away from the laboratory, diminishing the quality of administrative performance  within the
institution, and  limiting the individual's time to serve the community through external outreach
\cite{TenureBook,Tenure}. These momentum displacements can directly transform into negative productivity shocks to
  scientific output. As a result, there may be increased pressure for individuals in short-term contracts to produce
quantity over quality, which encourages the presentation of incomplete analysis  and diminishes the incentives to
perform sound science. These changing features may precipitate in a ``tragedy of the scientific commons.''

Aside from promoting circumspect research,  job security in academia diminishes the incentives for scientists to ``save
and store'' 
their knowledge for future liquidation in the case of employment emergency, and thus promotes the institution of ``open
science'' \cite{OpenScience}.  
However, a policy shift towards short-term contracts, along with the heightened value of intellectual property, may
alter the course of publicly funded ``open  science.'' This 
scientific commons emerged from the noble courts during the Renaissance as a hallmark of the scientific revolution and
now faces pressure from what has been termed ``intellectual capitalism,''  with the vast 
privatization of knowledge and innovation (``closed science'') occurring in  public universities and corporate R\&D
\cite{OpenScience}. An academic system that is dominated by short term contracts,  stymied by production incentives 
that favor quantity over quality, and jeopardized at the level of the ``open knowledge'' commons,  presents a new
institutional scenario revealing   selection pressures that could alter the birth and death rates of high-impact 
careers. 

  The purpose of this stochastic model is to  
 show how careers can become very susceptible to negative production shocks if the labor market is driven by  a
preferential capture 
 mechanism with $\gamma>1$ whereby early success of an individual can lead to  future advantage. However, this model
also shows that the onset of a fluctuation-dominant (volatile) labor market can also be amplified when the labor market
is governed by short-term contracts reinforced by a short-term appraisal system. In such a system, career sustainability
relies on continued recent short-term production, which can encourage rapid publication of low-quality science. In
professions where
 there is a high level of competition for employment, bottlenecks form whereby most careers stagnate and fail to rise
above an initial achievement barrier. Instead, these careers stagnate, and in a profession that shows no mercy for
production lulls, these careers undergo a ``sudden death'' because they were ``frozen out'' by a labor market that did
not provide insurance against endogenous fluctuations. Such a system is an employment ``death trap'' whereby most
careers stagnate and ``flat-line'' at zero production. However, at the same time, a small fraction of the population
overcomes the initial selection barrier and are championed as the ``big winners'', possibly only due to random chance. 
 
 Table \label{table:stats} demonstrates how the life expectancy decreases with increasing $c$ even for the linear
preferential capture model corresponding to $\pi =1$.
 With increasing $c$, the model simulates systems with shorter contracts (shorter appraisal ``memory'' timescales), and
so larger percentages of the population die before 
 characteristic ages $T_{c}(p)$, values  that decrease with increasing $c$ for a given $p$.
 
 \begin{table}
\begin{tabular}{@{\vrule height 10.5pt depth4pt  width0pt}lc|c|c|c|} &\multicolumn4c{ $T_{c}(p)$ as a \% of $T$, {\bf
(\% $T$)}}\\
\noalign{
\vskip-10pt}   \\
\cline{2-5}
\vrule depth 6pt width 0pt   & \ $p= 0.1\ $ & \  $p=0.25$ \ & \ $p=0.5$ \ & \ $p=0.75$ \\
\hline
\hline $c=0$ (long term) &  $0.94T$ & $0.98T$  & $1.00T$ & $1.00T$ \\
\hline
\hline $c=0.1$ &  $0.20T$ & $0.79T$  & $0.99T$ & $1.00T$ \\
\hline
\hline $c=1.0$ &  $0.01T$ & $0.02T$  & $0.05T$ & $0.15T$ \\
\hline
\hline $c=10.0$ (short term) &  $0.01T$ & $0.01T$  & $0.02T$ & $0.06T$ \\
\hline
\hline
\end{tabular}
\caption{  Decrease in  career life expectancy as a result of  short-term contract length in the $\pi =1$ linear
preferential capture model. The fraction $p$ of the population that experienced career termination before the 
crossover age $T_{c}(p)$: ``$p$ percent of the population died before reaching the age $L=T_{c}(p)$.'' 
As $c$ increases (recall the appraisal ``memory'' timescale is $1/c$) towards a short-term contract scenario, a
significant fraction of the population (increasing $p$) dies before reaching a smaller and smaller $T_{c}(p)$. The
empirical value of $T_{c}(p)$ is given as a percentage of the maximum career length $T$ corresponding to the stopping
time of the Monte Carlo simulation. The value  $T_{c}(p)$ is calculated using the equality $p = CDF(T<T_{c}(p))$, where
$CDF(T<L)$ is the cumulative distribution function of career length $L$. To estimate $CDF(T<L)$, we combine an ensemble
of 10 MC simulations for each $c$ value. In the model simulations we use $T\equiv 100$ periods. }
\label{table:stats}
\end{table}

\begin{figure}
\centering{\includegraphics[width=0.6\textwidth]{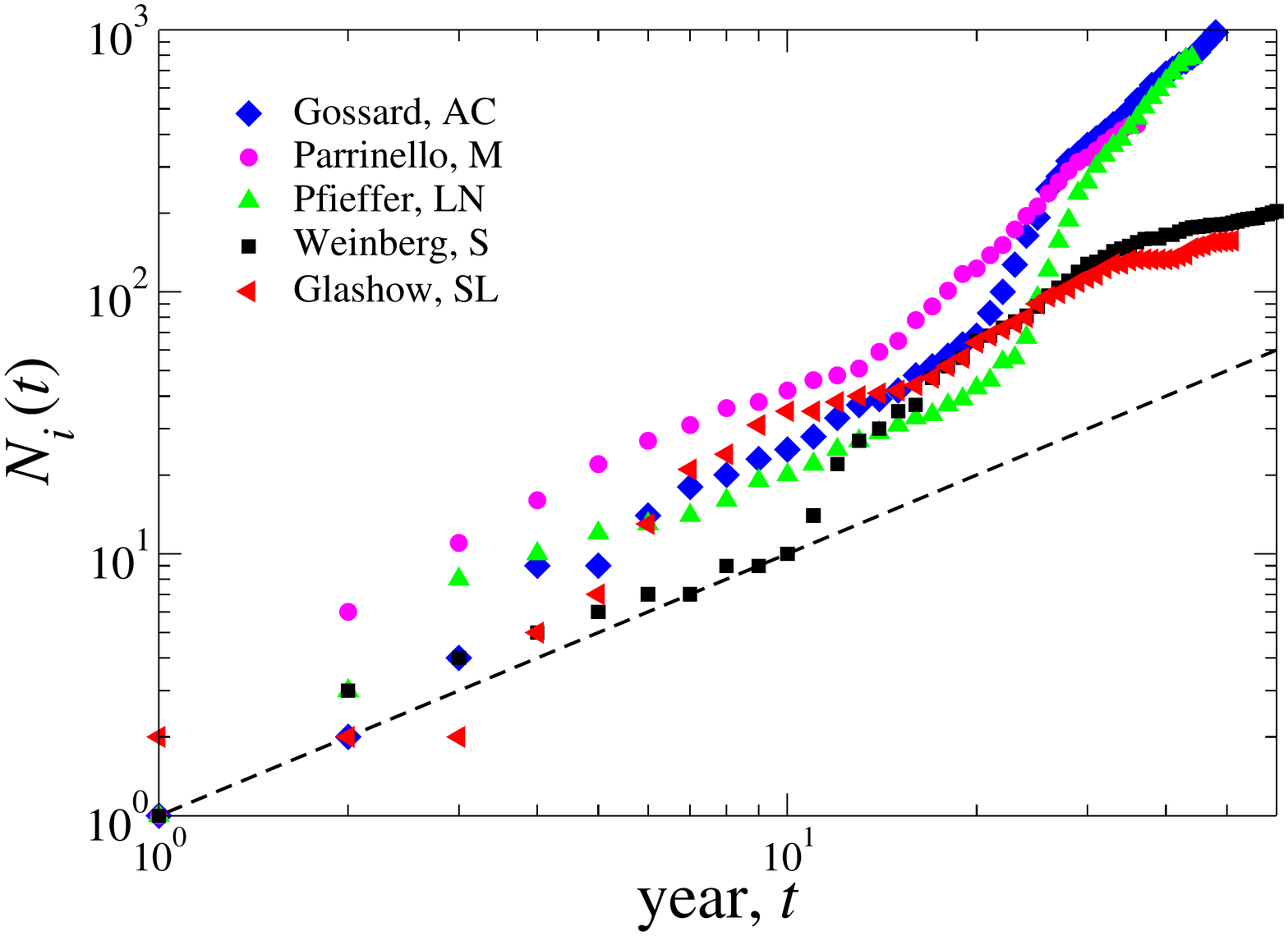}}
  \caption{ Positive career shocks likely associated with reputation boosts. Examples of career production
trajectories $N_{i}(t)$ that have significant deviations from the scaling
hypothesis in Eq. \ref{powlawN}. These significant deviations    likely follow extraordinary scientific discoveries (and
the publicity and reputation that are typically rewarded) which can vault a career and result in lasting benefits to the
individual. 
  }  \label{NTishocks}
\end{figure}

\begin{figure}
\centering{\includegraphics[width=0.9\textwidth]{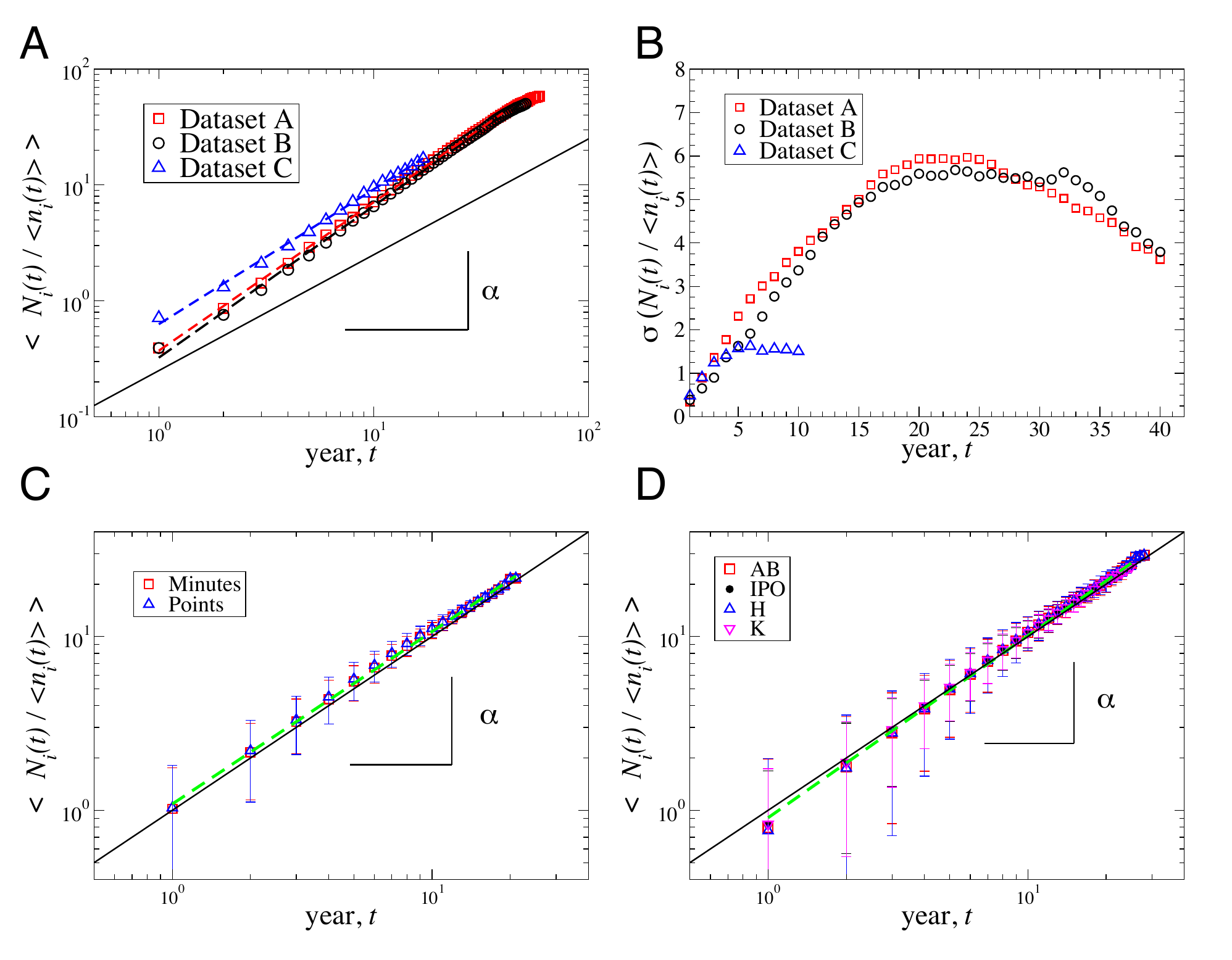}}
  \caption{ Regularities in the career trajectory $N_{i}(t)$. We analyze the normalized career trajectory
$N'_{i}(t)
\equiv N_{i}(t) / \langle n_{i} \rangle$ which allows us to aggregate $N_{i}(t)$ with varying publication rates $
\langle n_{i} \rangle$. As a result, we can better quantify the scaling exponent $\overline \alpha$ which quantifies the
acceleration of the typical career over time. We calculate $\overline \alpha$ using OLS regression on log-log scale of
the average normalized career trajectory $ \langle N'(t) \rangle \equiv \Big \langle \frac{N_{i}(t)}{ \langle n_{i}
\rangle} \Big \rangle$. For reference, each $N'_{i}(t)$ trajectory in panels A, B, and C has a corresponding best-fit
curve that is a dashed line.
  {\bf (A)} For the scientific careers, we calculate  $\overline \alpha$ values: $1.28\pm0.01$ for Dataset A, $1.31\pm0.01$
for Dataset B, and $1.15\pm0.02$ for Dataset C.
   These values are all significantly greater than unity, $\overline \alpha >1$, indicative of a systematic cumulative
advantage effect in science. {\bf (B)} 
   The standard deviation $\sigma N'(t)$ has a broad peak, likely related to career shocks that can significantly alter
the career trajectory.
   {\bf (C)} The average normalized career trajectory for NBA careers has $\overline \alpha \approx 1$ {\bf (D)} The average
normalized career trajectory for MLB careers has $\overline \alpha \approx 1$.  For visual comparison, the solid
straight black line in panels A,B and C correspond to a linear function with $\alpha =1$.}  \label{NT}
\end{figure}

\begin{figure}
\centering{\includegraphics[width=0.45\textwidth]{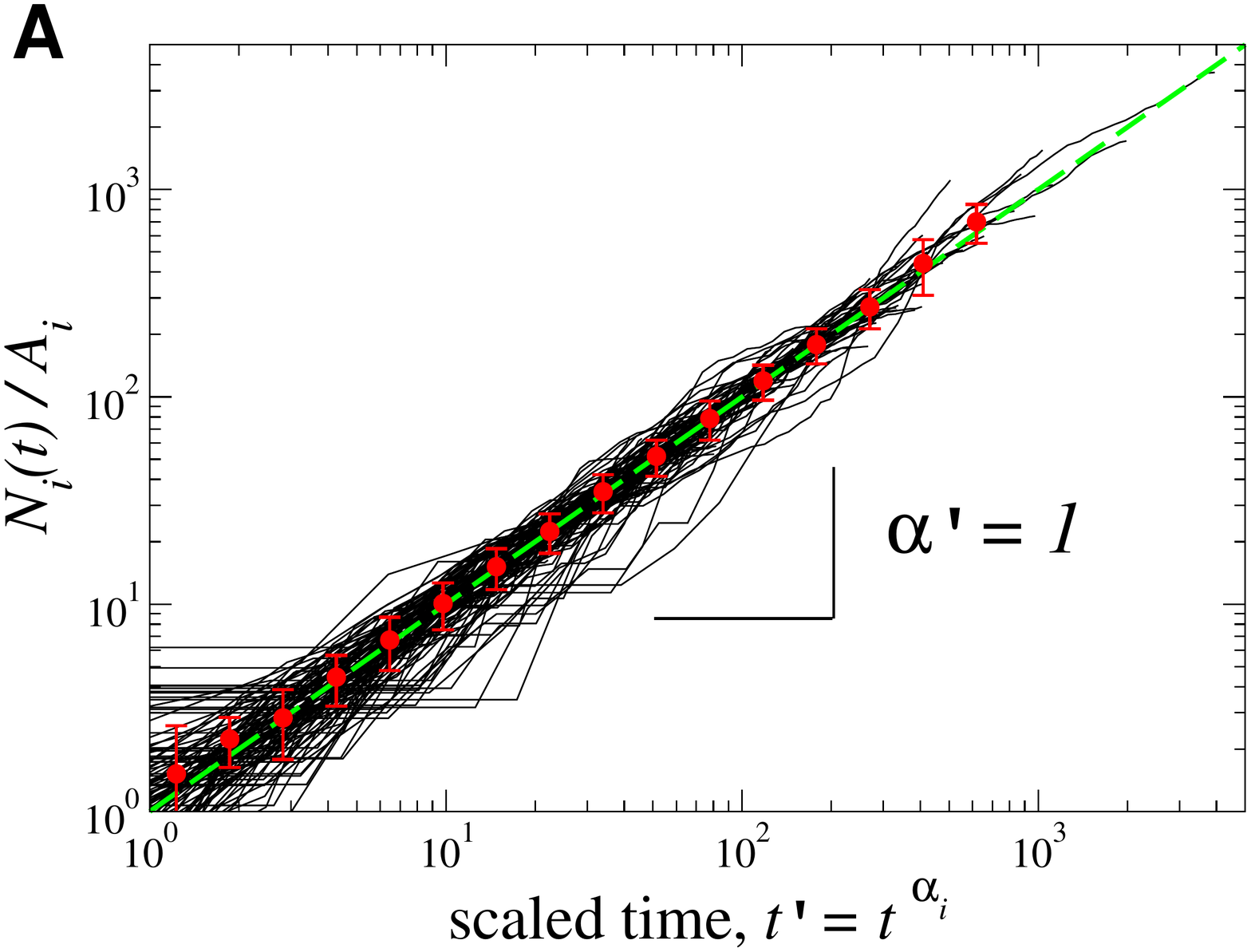}}
\centering{\includegraphics[width=0.45\textwidth]{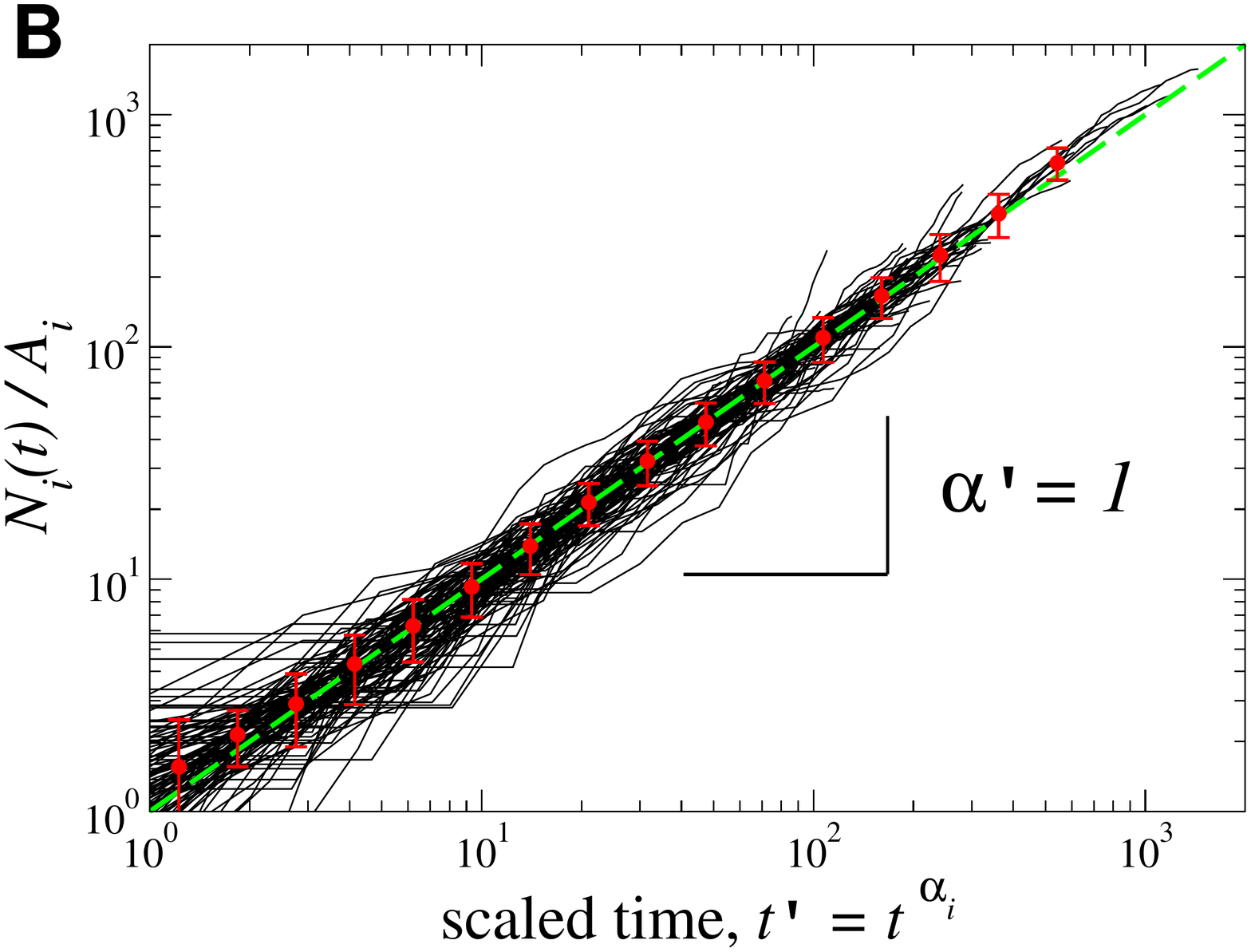}}\\
\centering{\includegraphics[width=0.45\textwidth]{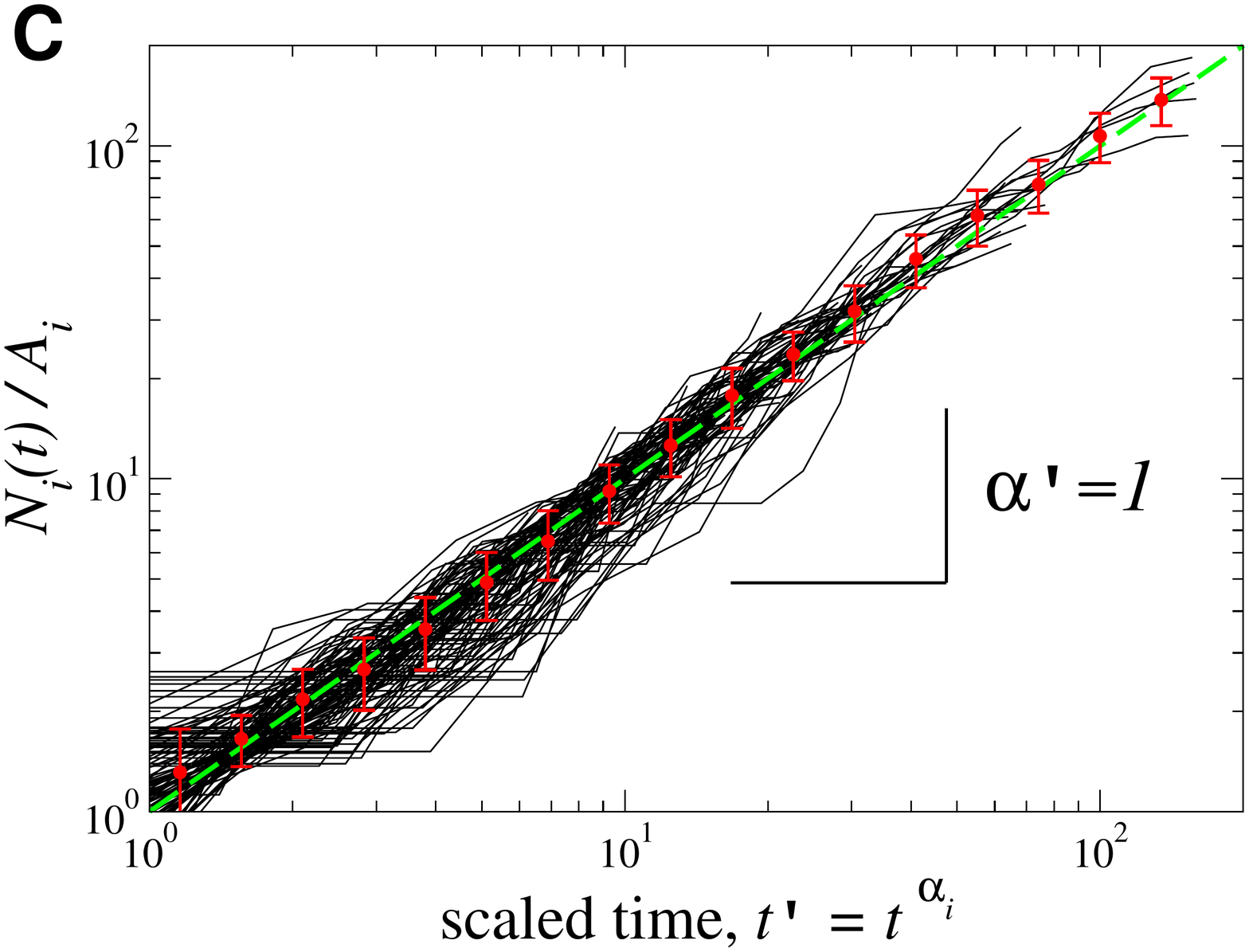}}
 \caption{ Using scaling methods to show approximate data collapse of each $N_{i}(t)$.
Normalized trajectory $\tilde N_{i}(t)
\equiv N_{i}(t)/A_{i}$ plotted using the scaled time $t'
\equiv
t^{\alpha_{i}}$ for each career over the time horizon $t \in [1,40]$ years.  We plot the 100 $\tilde N_{i}(t)$ curves
belonging to datasets [A], [B], and [C] in the corresponding panels. There is approximate data collapse  of all the
normalized trajectories $\tilde N_{i}(t)$ along the dashed green line corresponding to the rescaled career trajectory
$\tilde N_{i}(t) = t'$ with $\alpha' \equiv 1$ by construction. We also plot in red the corresponding 
average value $\langle \tilde N_{i}(t)\rangle$  with 1$\sigma$ error bars for logarithmically spaced $t'$ intervals.
Deviations from $\langle \tilde N_{i}(t)\rangle$ are indicative of career shocks which can significantly alter the
career trajectory. }  \label{NtotalScaled}
\end{figure}

\begin{figure}
\centering{\includegraphics[width=0.6\textwidth]{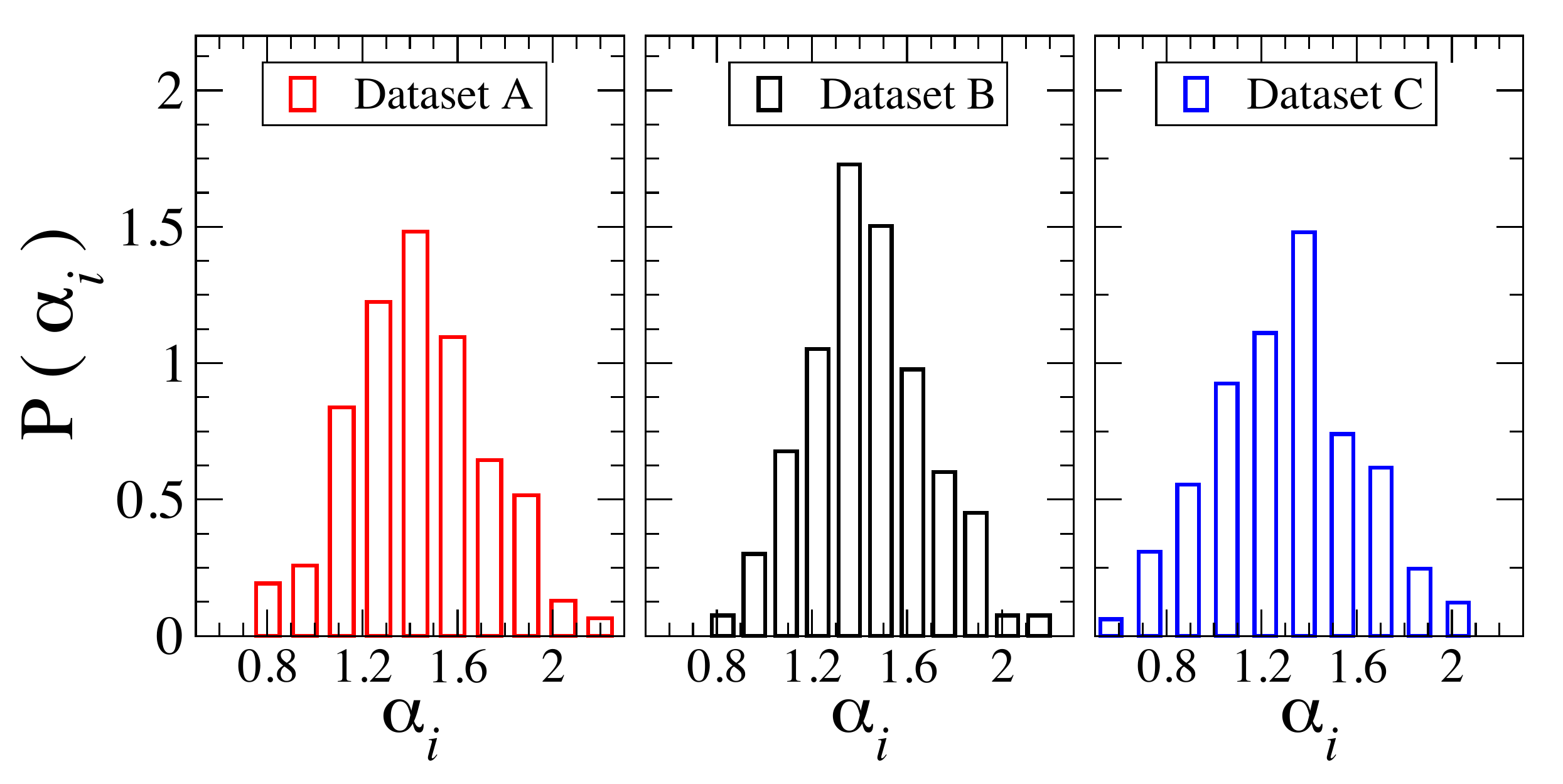}}
 \caption{ Increasing returns to scale $\alpha>1$. Probability distribution of the individual $\alpha_{i}$ values
calculated for each career using the scaling
model $N_{i}(t) \sim t^{\alpha_{i}}$ over time horizon $t \in [1,40]$ years. The average $\langle \alpha_{i} \rangle$
and standard deviation $\sigma(\alpha_{i})$ for each dataset are: $1.42 \pm 0.29$ [A], $1.44 \pm 0.26$ [B], $1.30 \pm
0.31$ [C].  The distribution of $\alpha_{i}$ values indicate that career trajectories are typically accelerating
$(\alpha_{i}>1)$, most likely the result of a cumulative advantage effect. }  \label{Alpha}
\end{figure}

\begin{figure*}
\centering{\includegraphics[width=0.85\textwidth]{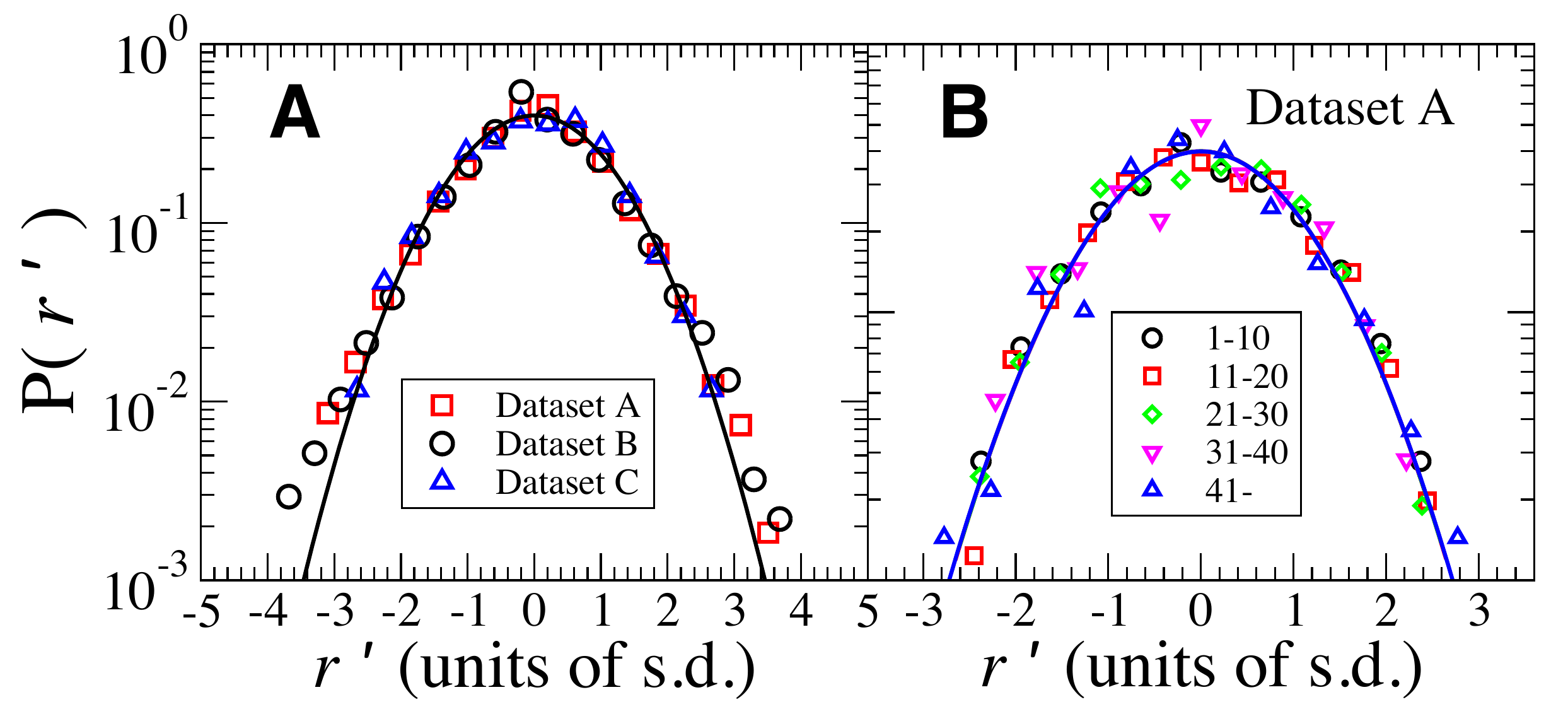}}
  \caption{ Universal patterns in underlying production fluctuations of scientists. Accounting for variable
individual publication factors, such as academic subfield or group collaboration size,
  we find that the normalized annual production change $r'_{i}(t) \equiv  [r_{i}(t)- \langle r \rangle_{i} ] /
\sigma_{i}$ is distributed 
  according to a Gaussian distribution, with $\langle r' \rangle =0$ and $\sigma(r')=1$ by construction (solid lines
show
   best-fit Guassian distributions using the maximum likelihood estimator method). This results indicates that the
Laplace distribution shown in Fig. \ref{dNPDF} results
  from a mixture of Gaussian distributions $P_{i}(r=\sigma_{i}r')$ that indicate that annual production is consistent with a proportional growth model.. } 
\label{drnormPDF}  
\end{figure*}

\begin{figure*}
\centering{\includegraphics[width=0.85\textwidth]{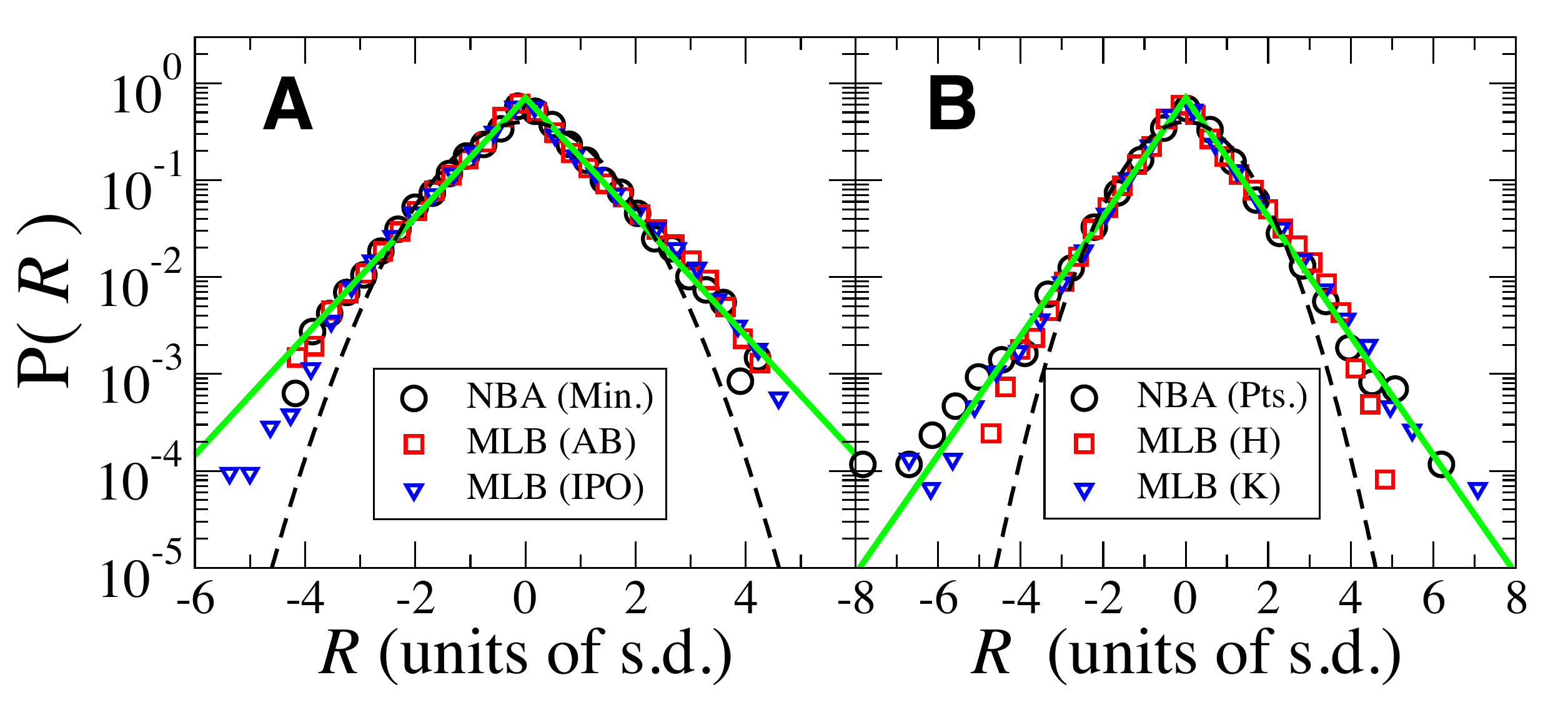}}
\centering{\includegraphics[width=0.85\textwidth]{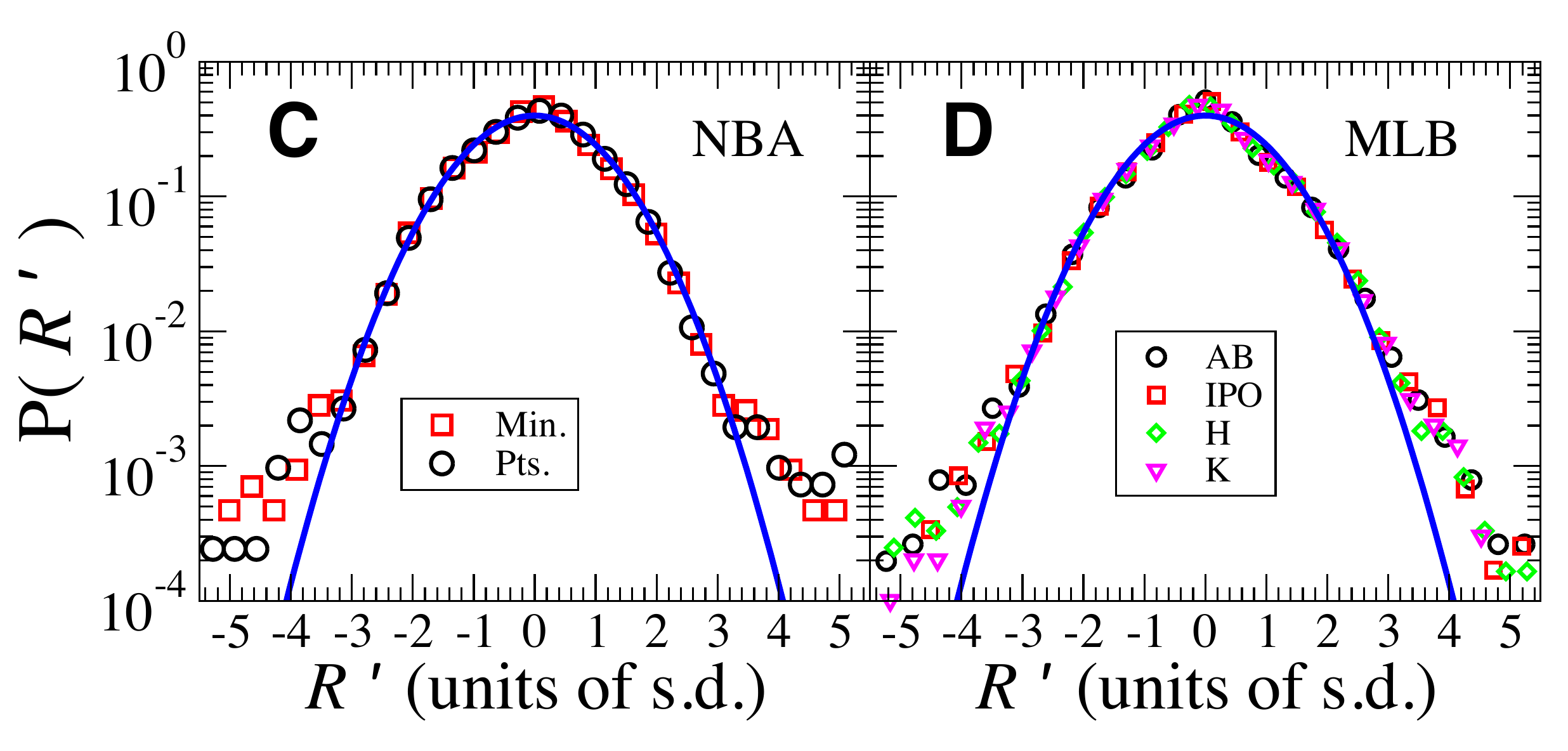}}
  \caption{  Universal patterns in the production fluctuations of athletes. 
For athlete careers  in the NBA and MLB we define production change for  (A,C) the change in the number of in-game
opportunities and (B,D)  the change in the number of in-game successes. (A,B) Since the detrended production change $R$ is
defined to have standard deviation $\sigma \equiv 1$, the  pdfs $P(R)$ approximately collapse onto a universal ``tent-shaped'' Laplace pdf (solid
green line). (C,D) For sports careers, we also
define a measure $R'$ which account for variable individual production factors,
such as propensity for injury, team position, etc. 
 As a result normalized annual growth rate $R'_{i} \equiv  [z_{i}(t)- \langle z(t) \rangle ] / \sigma_{z(t)}$ is normalized
twice, once to account for age factors and once to account for individual factors. The quantity $z_{i}(t) \equiv (r_{i}(t)-\langle r_{i} \rangle)/\sigma_{i}$ is normalized with respect to
individual factors, where $\langle r_{i} \rangle$ and $\sigma_{i}$ are the average and standard deviation of the
production change of career $i$. Then, we aggregate all $z_{i}(t)$  values for a given career year $t$ in order to
calculate the average $\langle z(t) \rangle$ and standard deviation $\sigma_{z(t)}$ over all careers. The final quantity
$R'_{i}$ represents a normalized annual production change which is distributed
  in the bulk according to a Gaussian distribution, with $\langle R' \rangle \approx 0$ and $\sigma(r') \approx1$ by
construction (solid lines show
   best-fit Guassian distributions using the maximum likelihood estimator method). This results indicates that the
tent-shaped distributions in (A,B)  results
  from a mixture of conditional Gaussian distributions $P_{i}(R=\sigma_{i}R')$ that indicate that annual production is consistent with a proportional growth model. } 
\label{drnormPDFsports}  
\end{figure*}

    \begin{figure}
\centering{\includegraphics[width=0.6\textwidth]{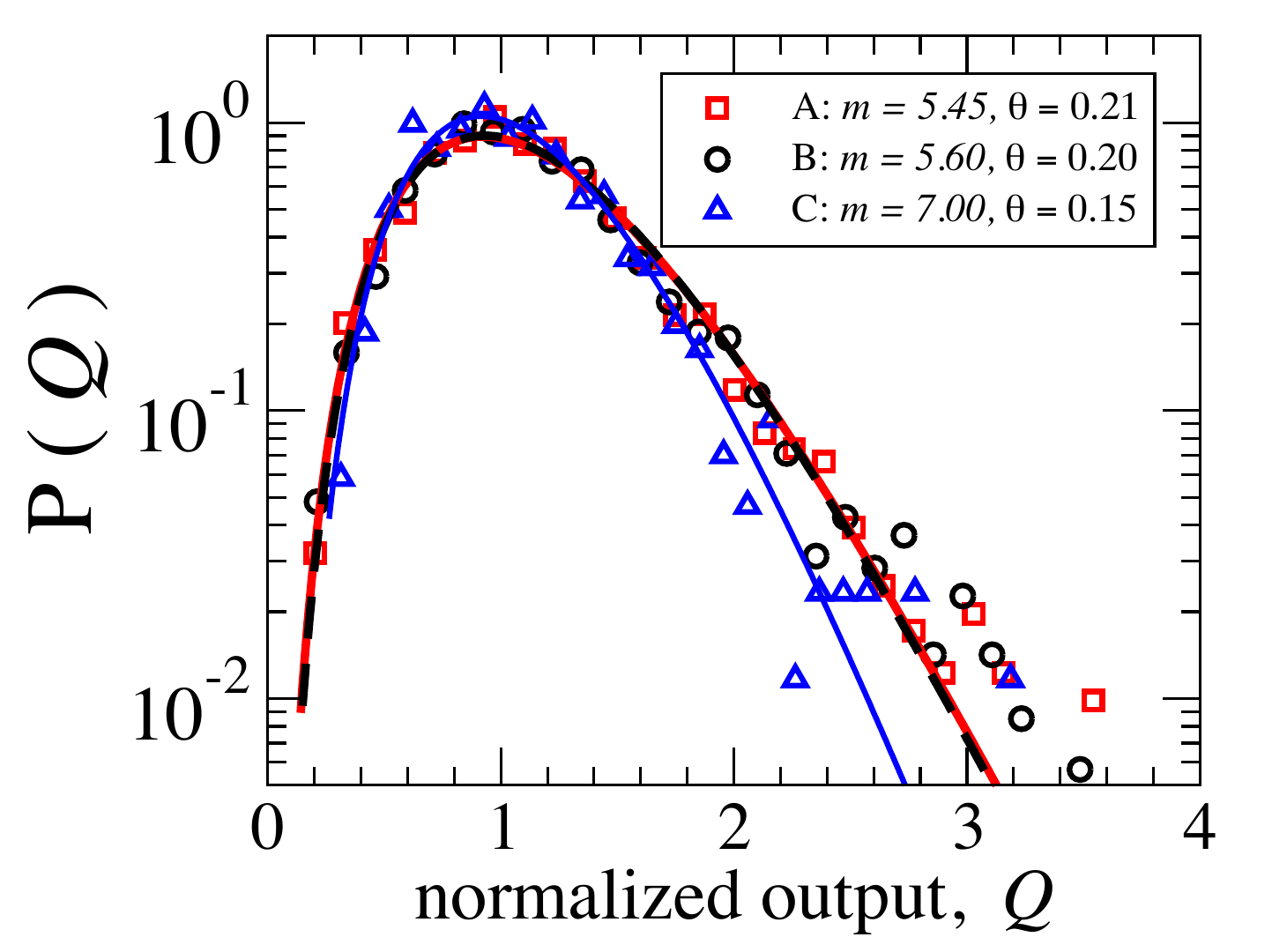}}
  \caption{ Universal micro-scale  output distribution $P(Q)$ which accounts for coauthorship variability. The
normalized 
  output $Q \propto n_{i}/k_{i}^{\gamma_{i}}$ is a residual output after we quantitatively account for the collaboration
size $k_{i}$ corresponding
   to the  number of distinct coauthors of author $i$. Each pdf is well-approximated by the Gamma distribution $P(Q)
\propto Q^{m-1}\exp[-Q/\theta]$  
   which suggests that production at the micro scale is governed by a Gamma L\'evy process.  We calculate the Gamma
distribution parameters using the
    maximum likelihood estimator method (distributions shown by solid and dashed curves), and find an insignificant
difference between [A] and [B] 
    scientists with Gamma shape  parameter $m$ and scale parameter $\theta$. However, for dataset [C] scientists, the 
output distribution is more
     skewed towards smaller $Q$ values, possibly reflecting the  relative advantage that senior scientists gain due to
reputation, experience, and
      knowledge spillover factors.    \label{ScalednPDF}  }
\end{figure}

  \begin{figure*}
\centering{\includegraphics[width=0.95\textwidth]{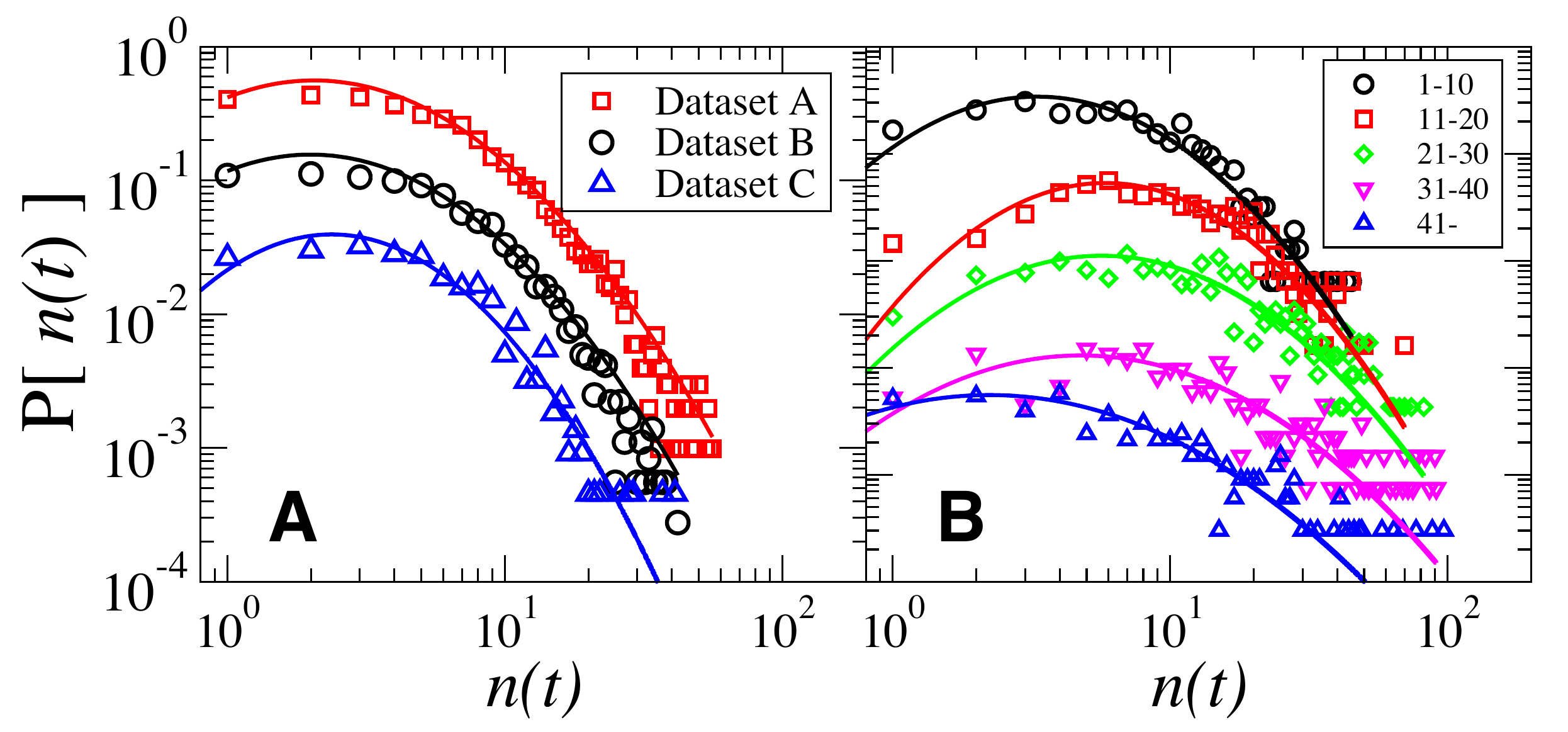}}
%\centering{\includegraphics[width=0.5\textwidth]{Figures/NSCALED_PDF_Asst100_Rand100_Top100.pdf}}
  \caption{ Aggregate production distributions can be deceiving. Unconditional distribution of annual publication
rate $n(t)$ appears as log-normal
 distributions because it is a mixture of 
  underlying distributions that depend strongly on collaboration factors. We define $n_{i}(t)$ as the number of papers
published in {\bf (A)} $\Delta t = 1$  and {\bf (B)} $\Delta t=2$ year periods, which reduces the finite-size effects arising from
the calendar  year labeling of publication dates. (A) We combine $n_{i}(t)$ values for all values of $t$, and find
excellent agreement between the empirical 
$P(n(t))$ data points and the log-normal model.
%the predicted log-normal $P[n(t)]$ shown as solid curves and defined in Eq.~[\ref{lognorm}], where 
 We use the maximum likelihood estimator method to calculate the log-normal parameters $\sigma_{L} \equiv \sigma(
\ln n )$ and $\mu = \langle \ln n \rangle$. {\bf (B)} In order to analyze the time-dependence of $P(n(t))$, we separate 
$n_{i}(t)$ values from Dataset A into 5 subsets, depending on the range $t$ years into the career, as indicated in the
figure legend. We offset each pdf by a constant factor in order to distinguish each pdf, which are also
well-approximated by   log-normal distributions (shown as solid curves). }  \label{nPDF}  
\end{figure*}

\begin{figure*}
\centering{\includegraphics[width=0.95\textwidth]{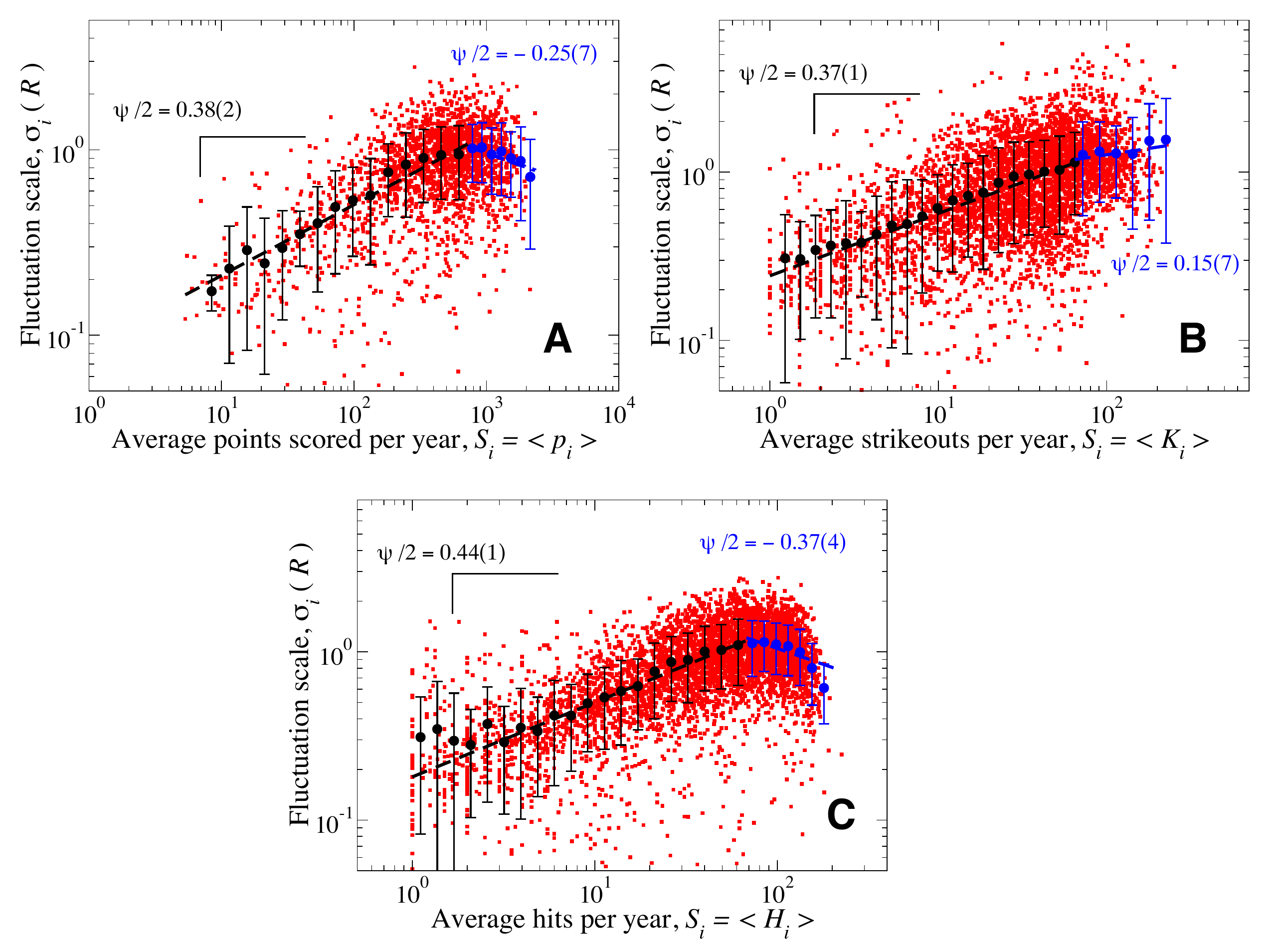}}
 \caption{ Quantifying the growth fluctuations of sports careers. The size variance relation for sports careers is similar
to academic careers for small $S_{i}$. 
However, for relatively large $S_{i}$ the relation becomes decreasing corresponding to $\psi <0$, analogous to what is
found for firm growth \cite{Growth1,Growth11,Growth13,Growth6}. The decreasing relation for $S_{i}> S_{c}$ likely
follows from the fact that in sports, there is a hard upper limit to the number of opportunities available to a player
in a given year. Hence, individuals with large $S_{i}$ are likely the starters on their teams, since it is neither
economical nor in the strategy of winning to keep players above a threshold value $S_{c}$ out of the game, and so these
players typically remain 
as positional starters except for episodic  leaves of absence due to injury. Hence, these players experience smaller
$\sigma_{i}(r)$ due to limitations to their potential for further career growth. However, players with $S_{i}< S_{c}$
are typically on the fringe of being released or provide alternative value to the team, and so these individuals
experience larger fluctuations in team play because they are easily dispensable, especially in a profession dominated by
short-contracts lasting sometimes less than a year. For each dataset, we use  careers  with career length $L_{i} \geq3$
seasons. {\bf (A)} NBA basketball players: Units of $\sigma_{i}(R)$ are normalized minutes played. We define the scaling
relation $\sigma_{i}(R) \sim \langle p_{i} \rangle^{\psi/2}$ between the average number of points scored per season $
 \langle p_{i} \rangle= \sum_{t=1}^{L_{i}}p_{i}(t)/L_{i}$ and the standard deviation $\sigma_{i}(R)$. In this way, we
utilize the average points per season as the proxy for the ability of a player to obtain future opportunities which are
realized as minutes played. Using $S_{c} \equiv 720$ points,  we calculate $\psi/2= 0.38 \pm 0.02$ (regression
coefficient $R=0.50$ and ANOVA F-test significance level $p \approx 0$) for $S_{i}<S_{c}$ and  $\psi/2= -0.25 \pm 0.07$
($R=0.15$ and  $p \approx 10^{-3}$)  for $S_{i}>S_{c}$. 
{\bf (B)} MLB pitchers: Units of $\sigma_{i}(R)$ are normalized IPO (innings pitched in outs). Interestingly,
$\sigma_{i}(R)$ continues to 
increase for $S_{i}>S_{c}$, possibly due to the relatively high career risk attributed to throwing  arm injury. Using
$S_{c} \equiv 65$ strikeouts,  we calculate $\psi/2= 0.37 \pm 0.01$ ($R=0.48$ and  $p \approx 0$) for $S_{i}<S_{c}$ and 
$\psi/2= +0.15 \pm 0.07$ ($R=0.07$ and  $p \approx 0.02$) for $S_{i}>S_{c}$. {\bf (C)} MLB batters: Units of
$\sigma_{i}(R)$ are normalized AB (at bats). Using $S_{c} \equiv 68$ hits, we calculate $\psi/2= 0.44 \pm 0.01$
($R=0.59$ and  $p \approx 0$) for $S_{i}<S_{c}$ and  $\psi/2= -0.37 \pm 0.03$ ($R=0.21$ and  $p \approx 0$)  for
$S_{i}>S_{c}$. The dashed black (blue) line in each panel is a least squares linear regression on log-log scale for all
data values with $S_{i}$ less (greater) than $S_{c}$. The  data shown with error bars   represent the average $\langle
\sigma_{i}(R) \rangle$ and corresponding 1 standard deviation  values calculated using equally spaced $S_{i}$ bins on
the logarithmic scale. }  \label{ptsSDscaling}
\end{figure*}  

\clearpage
\newpage

\begin{figure*}
\centering{\includegraphics[width=0.6\textwidth]{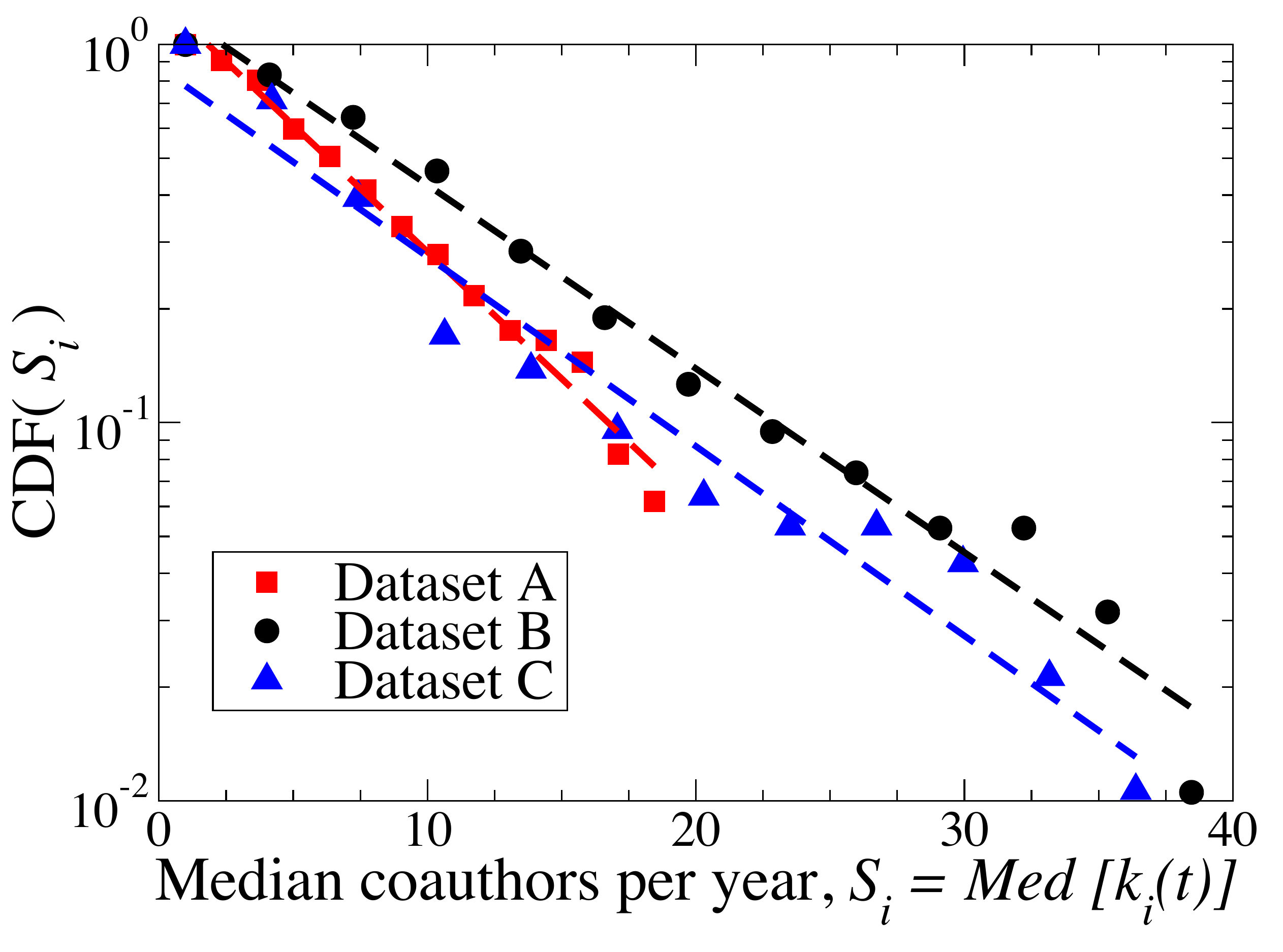}}
 \caption{  Exponential distributions of coauthor radius in Physics. We test the hypothesis that the distributions
$P(r)$ for annual production change
 $r$ (shown in Fig.
\ref{dNPDF}) follow 
 from an exponential mixing of Gaussians with varying fluctuation scale $\sigma_{i} \propto Med[ k_{i}(t) ]^{\psi/2}$.
An important criteria for this model is that the distribution of  $S_{i} \equiv Med[ k_{i}(t) ]$ is exponential, 
 $P(S_{i}) \sim \exp[-\lambda S_{i}]$. We plot the cumulative distribution function (CDF) $P(x> S_{i})$ for each
dataset, and confirm that the distributions are
 approximately linear on log-linear axes. Using linear regression, we calculate  $\lambda = 0.15 \pm 0.01$ [A], 
$\lambda = 0.11\pm 0.01$ [B], and $\lambda = 0.11 \pm 0.01$ [C].  
}  \label{NAvePapersCDF}
\end{figure*}

\begin{figure*}
\centering{\includegraphics[width=0.6\textwidth]{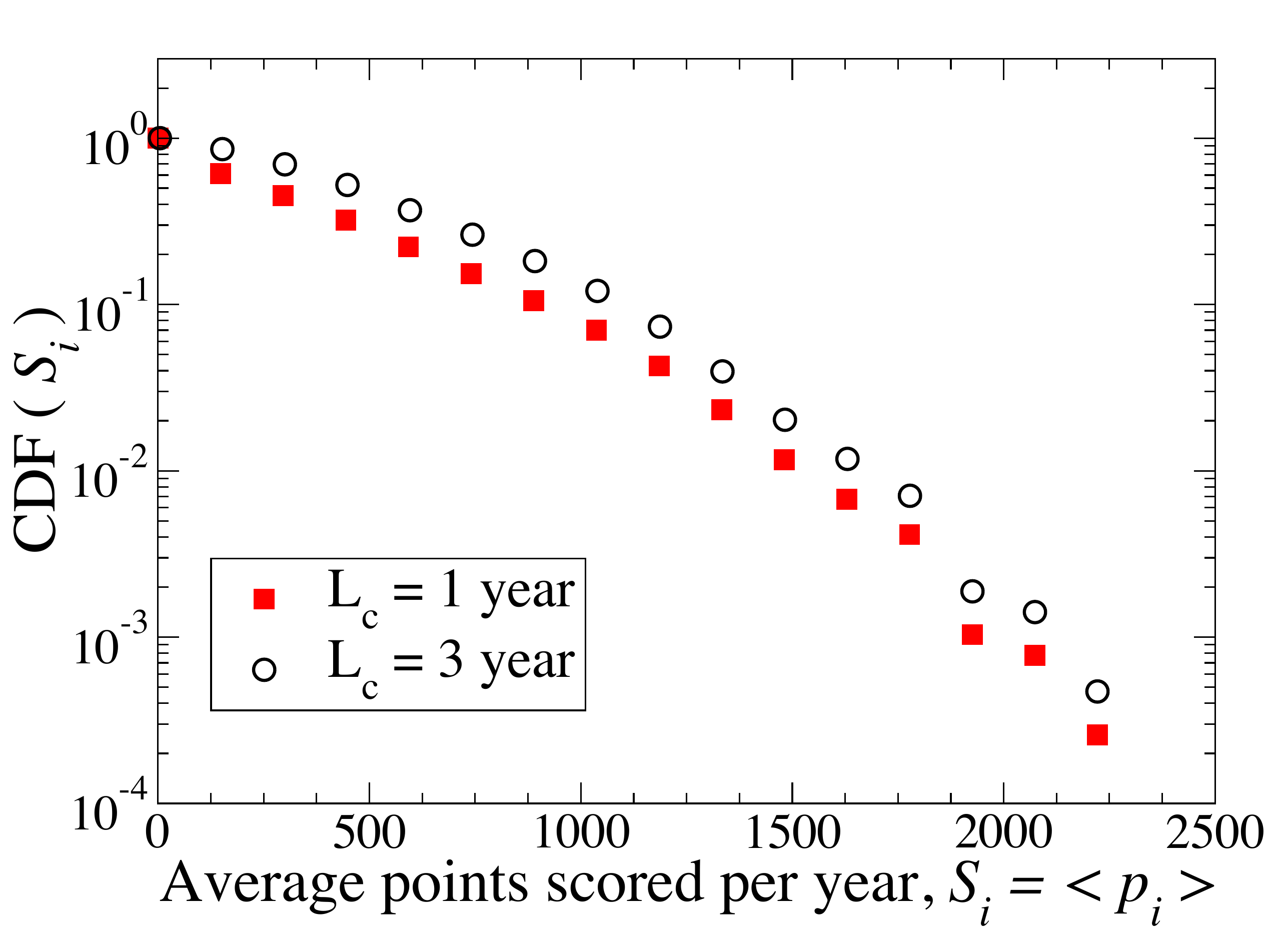}}
 \caption{  Approximately exponential distribution of scoring value in the NBA. We further test the hypothesis
that the distributions $P(R)$
 for annual production change
 $R$ in professional sports (shown in Fig. \ref{dNPDF} C and D) follow 
 from an exponential mixing of Gaussians with varying fluctuation scale $\sigma_{i} \propto \langle p_{i}
\rangle^{\psi/2}$. An important criteria for this model is that the distribution of ``team value'' $\langle p_{i}
\rangle$ is exponential, 
 $P(\langle p_{i} \rangle) \sim \exp[-\lambda \langle p_{i} \rangle]$. We plot the cumulative distribution function
(CDF) $P(x> \langle p_{i} \rangle)$ for each dataset, and confirm that the distributions are
 approximately linear on log-linear axes. We show the CDFs calculated using  all careers with career length $L_{i} \geq
L_{c}$ years, for $L_{c} = 1, 3$ years.} 
 \label{CDFpts}
\end{figure*}  

\clearpage
\newpage

\begin{figure*}
\centering{\includegraphics[width=0.85\textwidth]{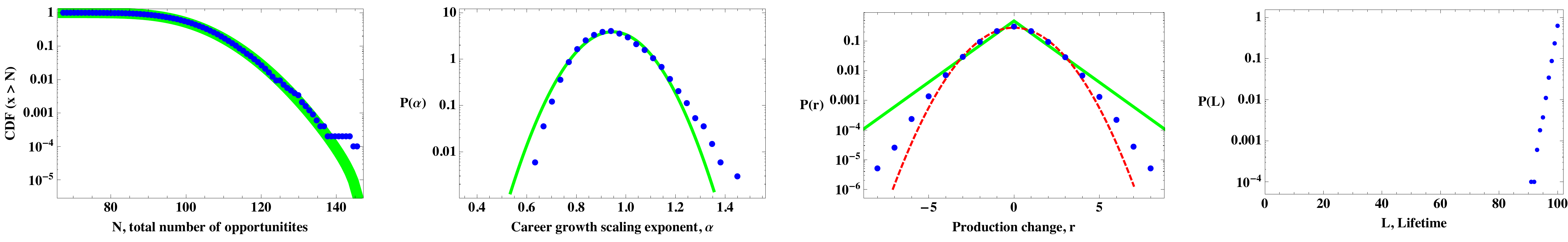}}
 \caption{  A production output null model with $\pi =0$ agrees with the predictions of a Poisson process. (Far
left) The cumulative distribution $CDF(x>N)$ is 
in excellent agreement with the  prediction of a  Poisson process with rate $\lambda_{p} =1$ and corresponding average $\langle N\rangle = \lambda_{p} T=100$. 
The solid green curve is the corresponding Poisson CDF using  $\langle N\rangle \equiv 100$.
(Middle left) Furthermore, the typical scaling exponent $\langle \alpha \rangle =1$ which is also consistent with
Poisson trajectories. (Middle right) The distribution of 
production changes is close to Gaussian. (Far right) The typical career length $L_{i}$ spans the entire system length
$T$, indicating low levels of career risk.} 
 \label{pi0C0}
\end{figure*}

\begin{figure*}
\centering{\includegraphics[width=0.99\textwidth, angle = 0]{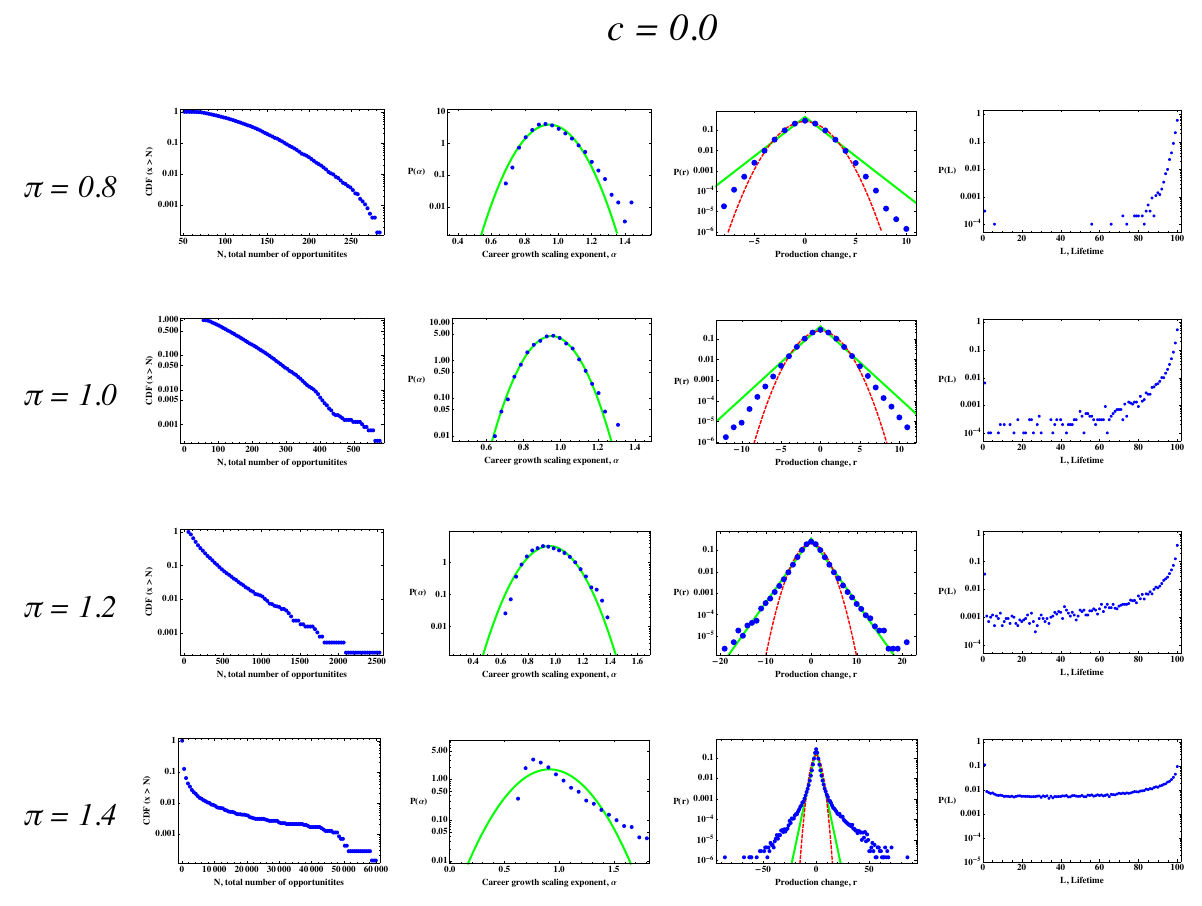}}
 \caption{ The production output model with $c =0$.  Results of  MC simulations for a ``long-term appraisal'' scenario. Careers are less vulnerable to low-production
 phases, and as a result, most agents sustain production throughout the career for a relatively large range of $\pi$
values.  } 
 \label{C0}
\end{figure*}  

\begin{figure*}
\centering{\includegraphics[width=0.99\textwidth, angle = 0]{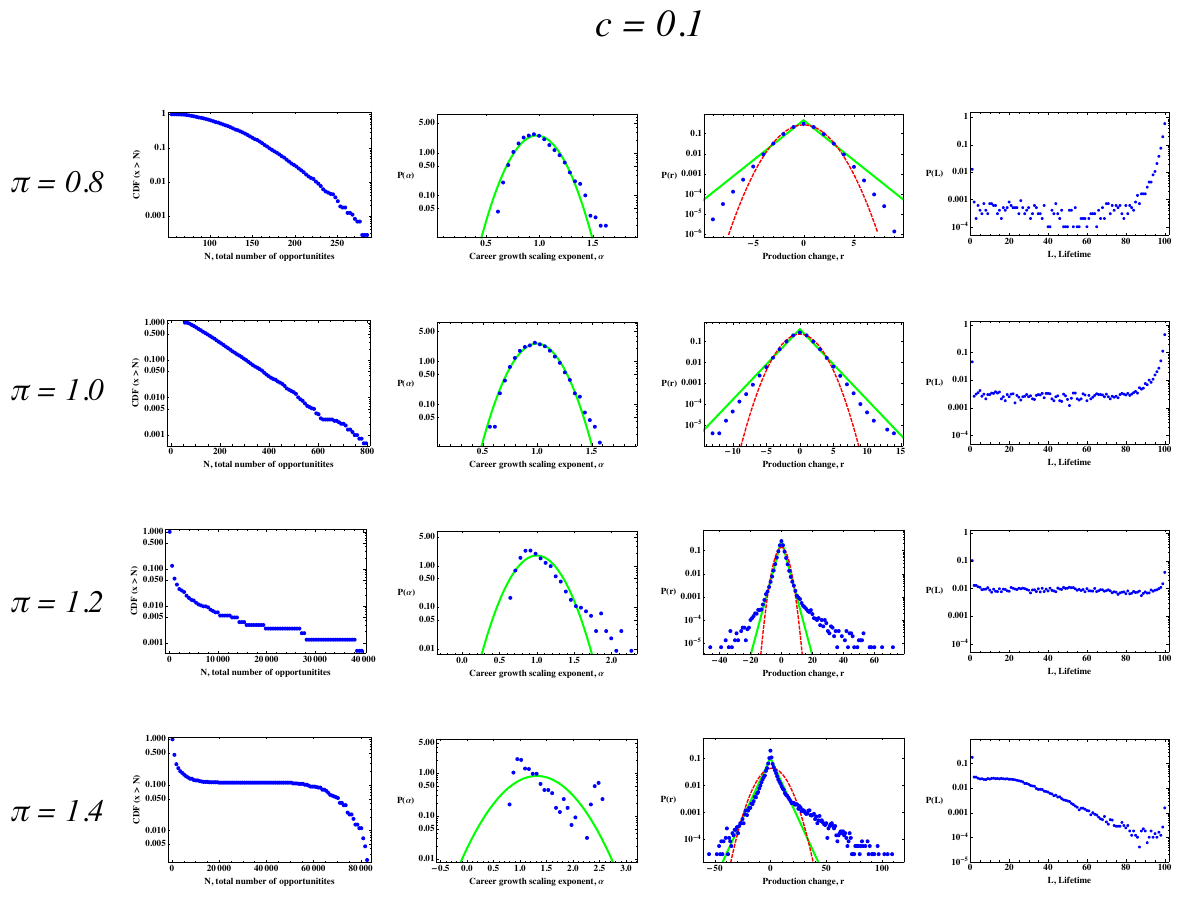}}
 \caption{   The production output model with $c =0.1$. Results of  MC simulations  for a ``medium-term appraisal'' scenario. The corresponding memory
time scale is approximately 10 time periods, and so only for significantly large $\pi =1.4$ do we observe a labor market
scenario in which there is a significant 
 death rate and just a few ``big winners'' corresponding to those agents with $\alpha \geq 1$.} 
 \label{C0p1}
\end{figure*}  

\begin{figure*}
\centering{\includegraphics[width=0.99\textwidth, angle = 0]{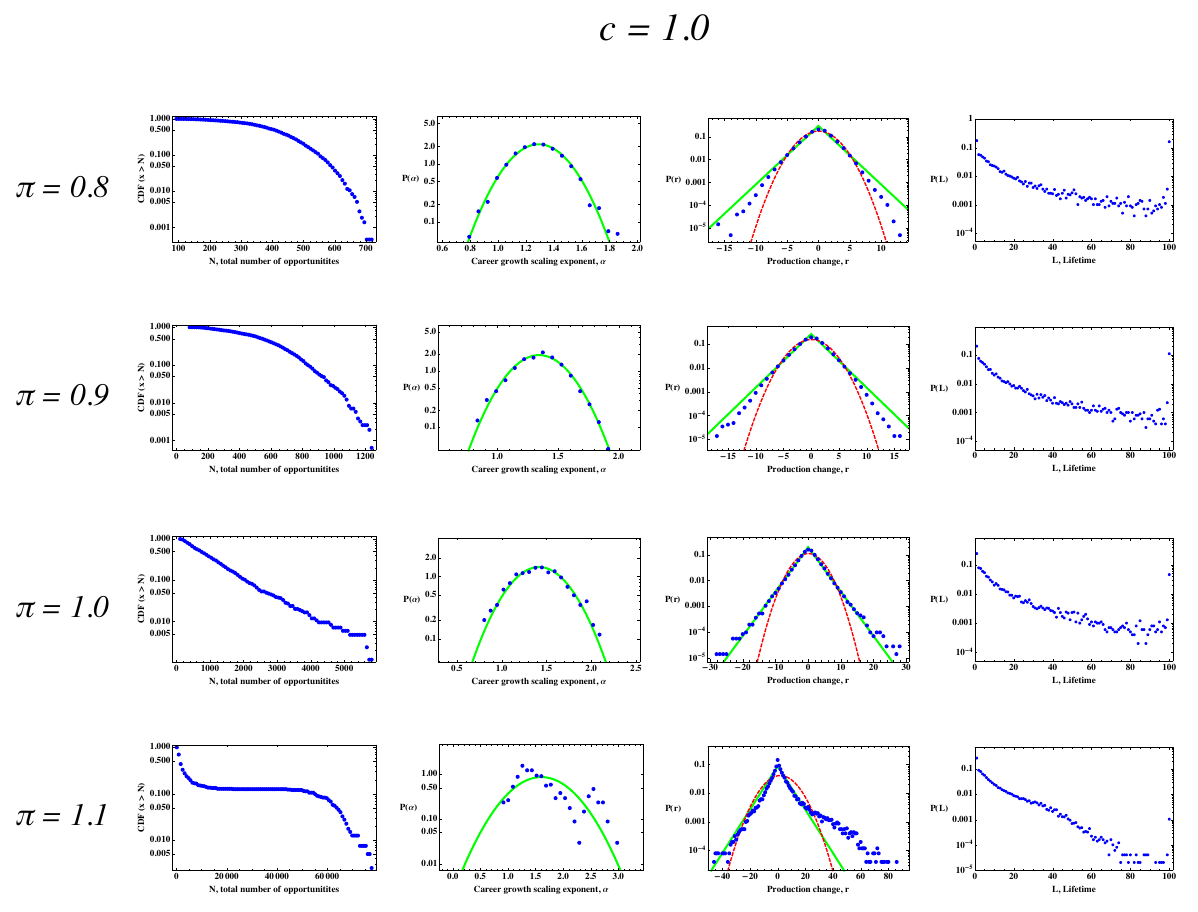}}
 \caption{   The production output model with $c =1.0$. Results of  MC simulations  for a ``short-term appraisal'' scenario. The corresponding memory time
scale is approximately 1 time period. Even for $\pi<1$, the system is driven by fluctuations that can cause career
``sudden death'' for a large fraction of the population. For $\pi >1$ we observe a very quick transition to a 
significant 
 death rate and just a few ``big winners'' corresponding to those agents with $\alpha \geq 1$.} 
 \label{C1p0}
\end{figure*}  

\begin{figure*}
\centering{\includegraphics[width=0.99\textwidth, angle = 0]{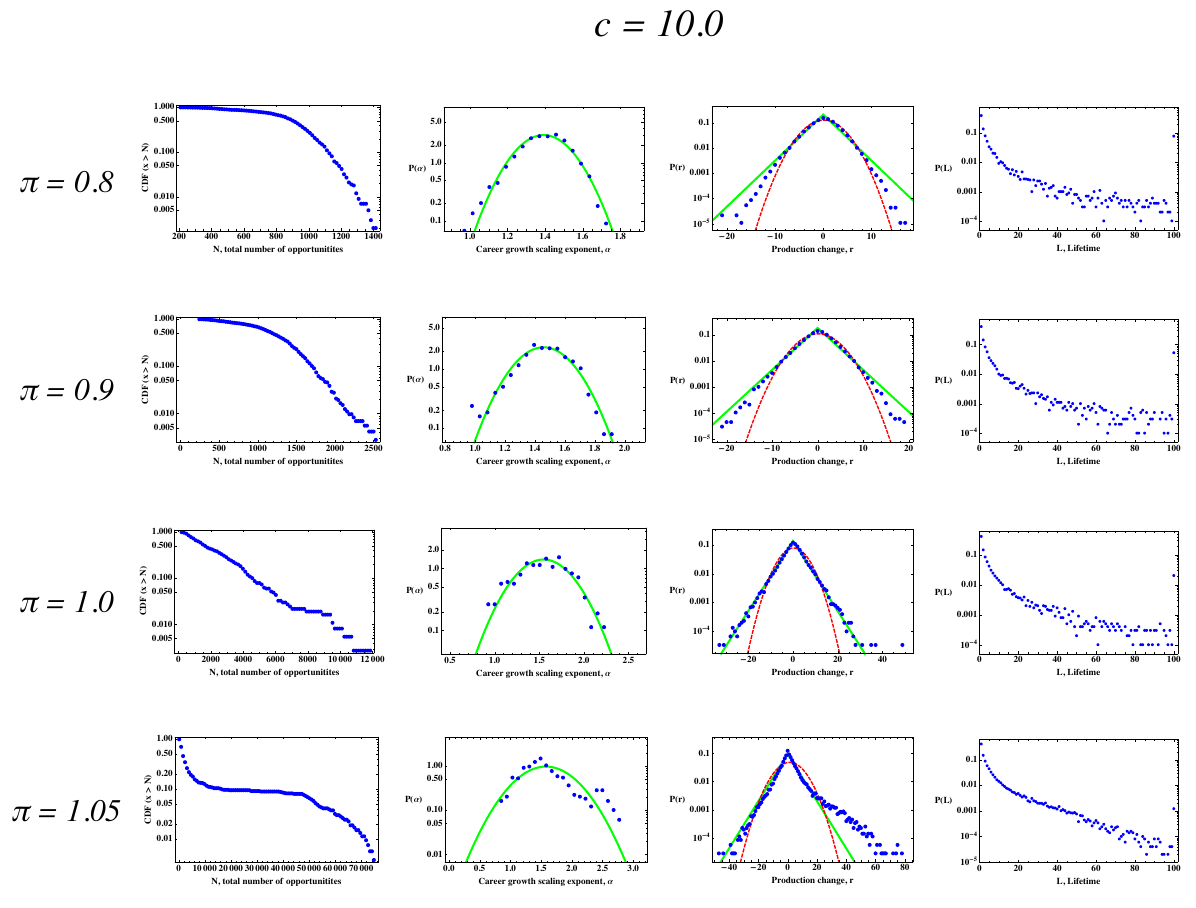}}
 \caption{   The production output model with $c =10.0$. Results of  MC simulations  for a ``zero-memory appraisal''
scenario wherein only the previous period
matters, $w_{i}(t) = n_{i}(t-1)$. Even for linear preferential capture $\pi =1$, the systems shows ``no mercy'' for
careers that are stagnant for possibly just one period. As a result, just a few ``lucky'' agents are able to survive the
initial fluctuations and end up dominating the system. For $\pi$ values close to unity, $\pi \rightarrow 1$, the systems
quickly becomes an employment ``death trap'' whereby most careers stagnate and ``flat-line.'' } 
 \label{C10p0}
\end{figure*}

\end{widetext}

\end{document}